\documentclass[%
 reprint,
 amsmath,amssymb,
 aps,
 prb,
]{revtex4-1}

\usepackage{graphicx}
\usepackage{dcolumn}
\usepackage{bm}

\usepackage{xcolor}
\usepackage{booktabs}
\newcommand{\Hsh}{$H_{sh}$}
\definecolor{darkBlue}{rgb}{0,0,0.6}

\newcommand{\kappaNb}{\ensuremath{\kappa_\mathrm{Nb}}}
\newcommand{\kappaNbSb}{\ensuremath{\kappa_\mathrm{Nb3Sn}}}
\begin{document}


\title{Analysis of Magnetic Vortex Dissipation in Sn-Segregated Boundaries in Nb$_3$Sn Superconducting RF Cavities}

\author{Jared Carlson}
\thanks{These authors contributed equally} 

\author{Alden Pack}%
 \author{Mark K.~Transtrum}%
 \email{mktranstrum@byu.edu}

\affiliation{Department of Physics and Astronomy, Brigham Young University, Provo, Utah 84602, USA}

\author{Jaeyel Lee}
\thanks{These authors contributed equally}

\author{David N.~Seidman}
\affiliation{ Department of Materials Science and Engineering, Northwestern University, Evanston, Illinois 60201, USA}

\author{Danilo B.~Liarte}
\thanks{These authors contributed equally}

\author{Nathan Sitaraman}
\thanks{These authors contributed equally}

\author{Alen Senanian}
\author{Michelle Kelley}
\author{James P.~Sethna}
\author{Tomas Arias}
\affiliation{Cornell University, Ithaca, New York 14853, USA}

\author{Sam Posen}
\affiliation{Fermi National Accelerator laboratory, Batavia, Illinois 60510, USA}


\date{\today}

\begin{abstract}
  We study mechanisms of vortex nucleation in Nb$_3$Sn Superconducting RF (SRF) cavities using a combination of experimental, theoretical, and computational methods.
  Scanning transmission electron microscopy (STEM) image and energy dispersive spectroscopy (EDS) of some Nb$_3$Sn cavities show Sn segregation at grain boundaries in Nb$_3$Sn with Sn concentration as high as $\sim$35 at.\% and widths $\sim$3 nm in chemical composition.
  Using ab initio calculations, we estimate the effect excess tin has on the local superconducting properties of the material.
  We model Sn segregation as a lowering of the local critical temperature.
  We then use time-dependent Ginzburg-Landau theory to understand the role of segregation on magnetic vortex nucleation.  
  Our simulations indicate that the grain boundaries act as both nucleation sites for vortex penetration and pinning sites for vortices after nucleation.
  Depending on the magnitude of the applied field, vortices may remain pinned in the grain boundary or penetrate the grain itself.
  We estimate the superconducting losses due to vortices filling grain boundaries and compare with observed performance degradation with higher magnetic fields.
  We estimate that the quality factor may decrease by an order of magnitude ($10^{10}$ to $10^9$) at typical operating fields if 0.03\% of the grain boundaries actively nucleate vortices.
  We additionally estimate the volume that would need to be filled with vortices to match experimental observations of cavity heating.


\end{abstract}

\maketitle

\section{\label{sec:Introduction}introduction}

Superconducting Radio-Frequency (SRF) cavities are used in accelerators to transfer energy to beams of charged particles.
Oscillating electric fields induce magnetic fields parallel to the surface of the cavity.
However, for very large induced magnetic fields, the superconducting Meissner state is unstable, so that a fundamental limit to the cavity's performance  is the material's ability to expel large magnetic fields.
For type-II materials, complete flux expulsion is thermodynamically stable up to a lower-critical field, $H_{c1}$, and a mixed state characterized by superconducting vortices is stable for fields up to an upper-critical field, $H_{c2}$.
Thus, by limiting the fields on the walls of the SRF cavities, the superconductor can be kept in the flux-free Meissner state, so that surface dissipation is extremely small and quality factors $\sim10^{10}$ can be achieved.
For magnetic fields parallel to the cavity surface, the superconducting Meissner state can be maintained above the stability limit in a metastable state up to a limit (for an ideal surface) of the so-called superheating field \Hsh\cite{LiarteSet2017}.
\Hsh\ has been studied extensively by the condensed matter community, primarily in the context of Ginzburg-Landau theory at ideal interfaces\cite{transtrum2011superheating, chapman1995superheating, dolgert1996superheating,kramer1968stability}.
Because high-power SRF cavities routinely operate in the metastable regime\cite{posen2015radio}, there has been renewed interested by the condensed matter community in the behavior of superconductors in large magnetic fields.
Calculations extend results to the semi-classical theory of Eilenberger theory in both the clean\cite{catelani2008temperature} and dirty\cite{lin2012effect} limits and Time-Dependent Ginzburg Landau (TDGL) simulations that account for material\cite{pack2019role} and spatial inhomogeneities\cite{aladyshkin2001best, burlachkov1991bean,vodolazov2000effect,soininen1994stability}.
In this paper, we explore the role of grain boundaries (GBs) in SRF cavity performance motivated by experimental observations of inhomogeneities in real-world SRF cavities.
It is well-established that grain boundaries can strongly affect the superconducting properties of Nb$_3$Sn.
They are known to pin vortices and determine critical currents in Nb$_3$Sn wires\cite{scanlan1975flux,godeke2006review}.
Compositional variation within Nb$_3$Sn grain-boundaries on the scale of a few nanometers have been observed\cite{suenaga1983chemical,sandim2013grain}.
This study brings together the expertise of many areas of condensed matter and accelerator physics to explore fundamental physics of superconductors in extreme conditions and connect those results to real systems.

Recently there has been significant progress towards the employment of Nb$_3$Sn in SRF cavities as a higher performance alternative to the industry standard Nb for next generation particle accelerator applications \cite{posen2017nb3sn,posen2014advances}.
Nb$_3$Sn cavities are prepared with Nb$_3$Sn films $\sim$2 $\mu$m (nearly 20 penetration depths) thick coated on the surface of Nb cavities using the Sn vapor-diffusion process.
Nb$_3$Sn is an intermetallic alloy with A15 crystal structure; it
is a promising material for next-generation SRF cavities in large part because of its large (predicted) superheating field ($\sim$400 mT).
It also has a higher critical temperature (Tc $\sim$18K), making possible a higher quality factor (Q$_0$, another critical metric of cavity performance) at a given temperature compared to Nb (Tc $\sim$9K).

In practice, however, real world Nb$_3$Sn cavities quench well below the theoretically predicted value.
The maximum accelerating gradient that has been achieved within these cavities is about 24 MV/m, which corresponds to the surface magnetic field of $\sim$98 mT.
These cavities exhibit a high Q$_0$ of $\sim10^{10}$ at 4.2 K \cite{posen2017nb3sn,posen2019nb3sn}; however, in some cavities, Q$_0$ begins to degrade significantly before the limiting quench field is reached, a phenomenon described as ``Q-slope''\cite{muller2000status}.

Understanding the mechanism underlying the Q-slope phenomenon is an important question for cavity development.
Several mechanisms have been proposed in terms of imperfections in the Nb$_3$Sn coatings \cite{gurevich2017theory,hall2017new,posen2015radio} such as  thin grains \cite{lee2018atomic,trenikhina2017performance} and Sn-deficient regions \cite{becker2015analysis}. 
Another potential mechanism that may have detrimental effects on the performance of Nb$_3$Sn cavities is Sn segregation at grain boundaries \cite{lee2019grain}.
In some Nb$_3$Sn coatings, tin concentrations as high as $\sim$35 at.\% have been observed in grain boundaries with the segregated zone extending by as much as $\sim$3 nm, comparable to the coherence length of Nb$_3$Sn ($\sim$3 nm).
Because of the inferior superconducting properties, magnetic flux may penetrate the segregated region, degrade Q$_0$, and lead to premature quench.

\begin{figure}
    \includegraphics[width=\columnwidth]{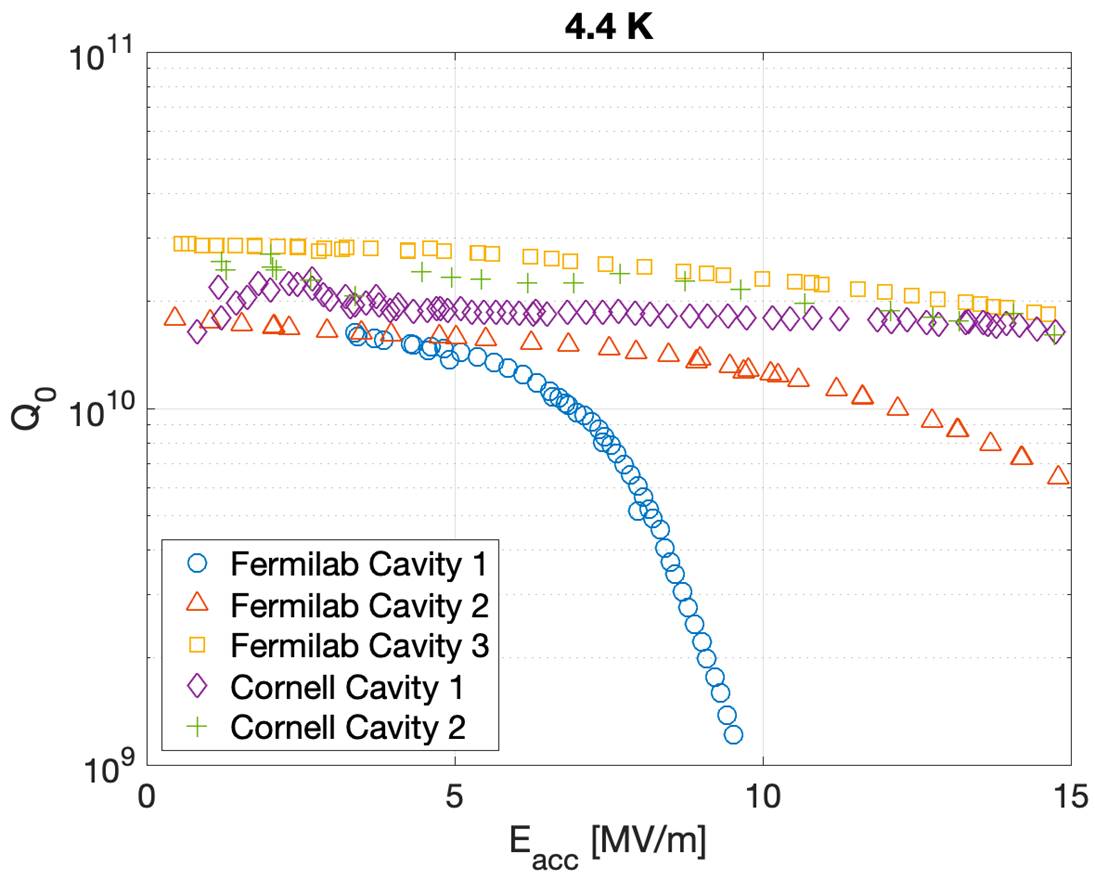}
    \caption{Q vs E curve of various Nb$_3$Sn SRF cavities coated at Fermilab and Cornell. The GBs of a witness sample (Fermilab Cavity 3) and direct cutout (Cornell Cavity 2) of a high-performance cavity are characterized in STEM-EDS and showed no Sn segregation at GBs. On the other hand, witness samples of Fermilab Cavity 1 and 2, which show Q-slope starting at ~8 MV/m, were found to have Sn segregation at GBs (reprint from \cite{lee2019grain}). }\label{fig:QvsE}
\end{figure}

In support of this hypothesis, witness samples coated with high-performance (quench at $\sim$24 MV/m with Q$\sim$10$^{10}$ at 4.4 K) Nb$_3$Sn cavities at Fermilab did not show Sn segregation at the grain boundaries in energy dispersive X-ray spectroscopy (EDS) and in scanning transmission electron microscopy (STEM).
Similarly, a direct cutout from a high-performance Nb$_3$Sn cavity fabricated at Cornell did not show Sn segregation at grain boundaries within the detection limit of STEM-EDS.
In contrast, Nb$_3$Sn cavities, which show Sn segregation at grain boundaries in witness samples
coated together with the cavities, displayed negative Q-slope for accelerating fields in the 5-15 MV/m range, see Fermilab cavity 1 and 2, Fig.~\ref{fig:QvsE}.
These experimental results, summarized in Fig.~\ref{fig:QvsE}, suggest a potential link between Sn segregation at grain boundaries and cavity performance \cite{lee2019grain}.

Experimentally, it is difficult to isolate the effects of Sn segregation at grain boundaries from other imperfections, such as Sn-deficient regions and surface roughness.
To overcome these challenges, we use numerical tools to theoretically understand the role of segregation in grain boundaries for SRF cavity performance.
We use density functional theory to estimate the effective $T_c$ of the material within the segregation zone.
Next, we use time-dependent Ginzburg-Landau simulations with spatial varying material properties motivated by the ab initio DFT calculations.
Time-dependent Ginzburg-Landau theory allows us to conduct numerical experiments on a mesoscale that probe the role of grain boundaries and segregated zones for vortex nucleation, pinning, and quenching.
Finally, motivated by our TDGL simulations, we estimate power dissipated by vortex nucleation within segregated grain boundaries during an RF cycle and make quantitative comparisons to actual SRF cavities.

This paper is organized as follows:
First, we present experimental and first-principles analyses of grain-boundary defect segregation in Nb$_3$Sn cavities in section~\ref{sec:Experimental}.
We then report on first principles DFT calculations of superconducting properties for segregation zones in section~\ref{sec:TcCalcs} and time-dependent Ginzburg-Landau simulations of flux penetration in section~\ref{sec:TDGL}.
We estimate the effect on cavity performance in section~\ref{sec:VortexTheory}.
Our numerical experiments isolate the effects of Sn-segregated grain boundaries from potentially confounding mechanisms.
Our results indicate that the effects of Sn-segregated grain boundaries alone are consistent with observed behaviors.
Specifically, grain boundaries may nucleate and then trap a limited number of vortices, leading to degradation in the cavity's quality factor.
We conclude by discussing the implications of these results for cavity development and further theoretical studies in section~\ref{sec:conclusion}.

\section{\label{sec:Experimental}Experimental and Theoretical Study of Nb$_3$Sn Defect Segregation}

\begin{figure}
    \includegraphics[width=2.5in]{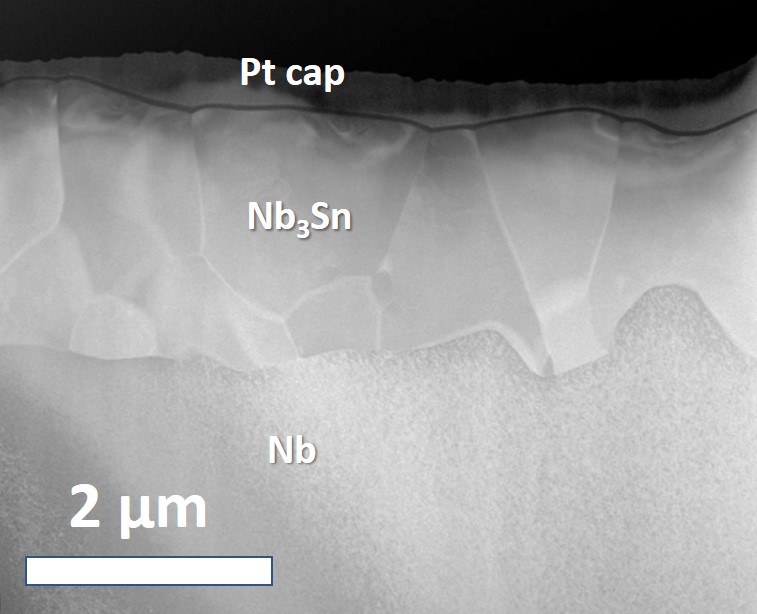}
    \caption{HAADF-STEM image of a typical $\sim$2 $\mu$m thick Nb$_3$Sn coating on Nb prepared by Sn vapor-diffusion. }\label{fig:STEMimageNb3Sn}
\end{figure}

\begin{figure}
    \includegraphics[width=2.5in]{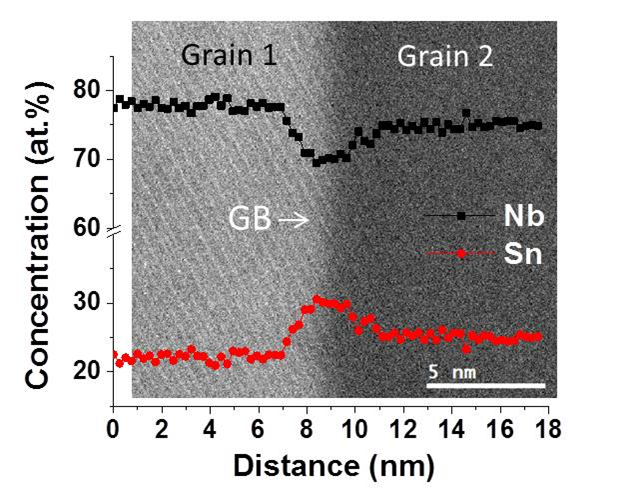}
    \caption{The HAADF-STEM image and corresponding Nb and Sn concentration profiles across the GB between Grain 1 and Grain 2. Sn is segregated at the GB up to $\sim$33 at.\% Sn and the width of the Sn segregated region is $\sim$3 nm.}\label{fig:GBsegregation}
\end{figure}

\begin{figure}
    \includegraphics[width=2.5in]{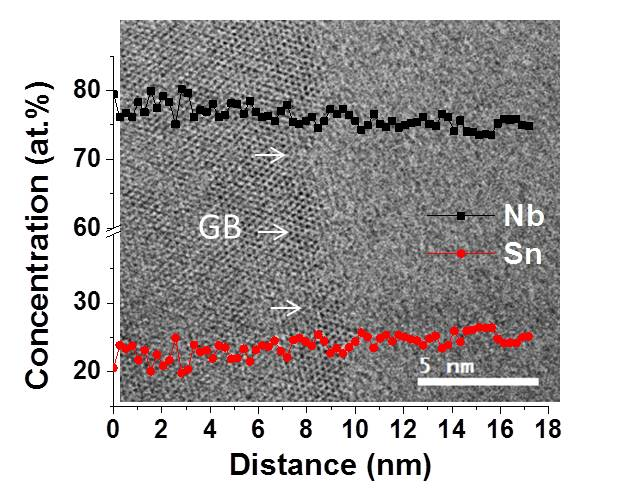}
    \caption{BF-STEM and corresponding Nb and Sn concentration profiles across a GB from a witness sample of high-performance Nb$_3$Sn SRF cavity prepared at Fermilab. }\label{fig:noGBsegregation}
\end{figure}

We have systematically investigated the composition of GBs in Nb$_3$Sn coatings and the origin of Sn-segregation at GBs using experimental and first-principles methods. The high angle annular dark field (HAADF)-STEM image in Figure \ref{fig:STEMimageNb3Sn} displays a Nb$_3$Sn coating on Nb prepared by the Sn vapor diffusion process using coating parameters from the early stage of the development of Nb$_3$Sn films at Fermilab \cite{posen2017nb3sn}. EDS mapping is performed across a GB in a Nb$_3$Sn coating prepared by the same coating parameters and reveals that Sn is segregated at the GB, Figure \ref{fig:GBsegregation}. The maximum concentration of Sn at the GB is $\sim$33 at.\% and the width of the Sn segregated region is $\sim$3 nm. The Gibbsian interfacial excess of Sn is $\sim$16 atom/nm$^2$. Previous analyses of the structures of Sn-segregated GBs in Nb$_3$Sn employing HR-STEM and first-principle calculations indicated that most of the segregated Sn exist as Sn-antisite defects near the GBs rather than forming Sn-rich phases such as Nb$_6$Sn$_5$ or other non-equilibrium phases \cite{lee2019grain, DILLON20076208}.  Another GB from a witness sample of a high-performance cavity, which is prepared by the coating procedure with less Sn and higher furnace temperature than the previous coating to mitigate the Sn-segregation at GBs \cite{lee2019grain}, is characterized by HR-STEM EDS, Fig. \ref{fig:noGBsegregation}. It is noted that there is no Sn segregation at the GB within the detection limit of EDS ($\sim$1 at.\%). This indicates a possible correlation between the Sn segregation at GBs and cavity performance, and motivates us to investigate the effects of Sn-segregation on GB properties using DFT and Ginzburg-Landau calculations.

To investigate the origin of the Sn-segregation in the Nb$_3$Sn coating, we consider our previous experimental study of Nb$_3$Sn GBs using STEM-EDS and APT analyses~\cite{lee2019grain}. Nine batches of coatings were characterized by randomly selecting 2-4 GBs as representatives for each coating and analyzing them. Our results show that the effect of the GB structure (characterized by five degrees of freedom given by the rotation axis (c), the disorientation angle ($\theta$), and the normal vector to the GB plane (n)) on Sn-segregation is insignificant, which is in contrast with the usual behavior of GB segregation behaviors of alloys in an equilibrium state \cite{seidman2002subnanoscale, rittner1997solute}. Even small-angle GB reveal a similar amount of Sn-segregation at GBs to high-angle GBs nearby. In parallel, we performed a detailed DFT study surveying a variety of GB structures in Nb$_3$Sn including both symmetric tilt and twist GBs, identifying a number of low energy structures \cite{kelley2020abinitio}.  We found a rapid decay of local strain while moving away from the GB, and a disruption to the electronic structure that decays to bulk behavior $\sim$1--1.5 nm from the GB. Importantly, this disruption reduces the Fermi-level density-of-states by more than a factor of 2 in the core region. Associated with this density-of-states reduction, we also found a strong binding of tin on niobium antisite defects to GBs that extends outward to the same distances of $\sim$1--1.5 nm.

Based on these results, we consider both kinetic and thermodynamic factors that may contribute to Sn-segregation at GBs. For the kinetic factor, we considered the Sn-diffusion process along GBs, and it is found that the amount of Sn-segregation at GBs in Nb$_3$Sn is affected by Sn-flux during the coating process: higher Sn-flux causes a more significant amount of Sn-segregation at GBs. This suggests that the kinetic Sn GB diffusion process plays a significant role in determining the chemical composition of GB. For the thermodynamic factor, the attractive electronic interaction between Sn-antisite defects and GBs may also play an important role, as we find that this interaction has a range similar to the observed radius of Sn-segregation, and depends only weakly on the structure of the GB. The current coating procedures at Fermilab and Cornell successfully mitigates the Sn-segregation at GBs by annealing without the external Sn-source \cite{kempshall2002grain}.

\begin{figure}
    \includegraphics[width=2.5in]{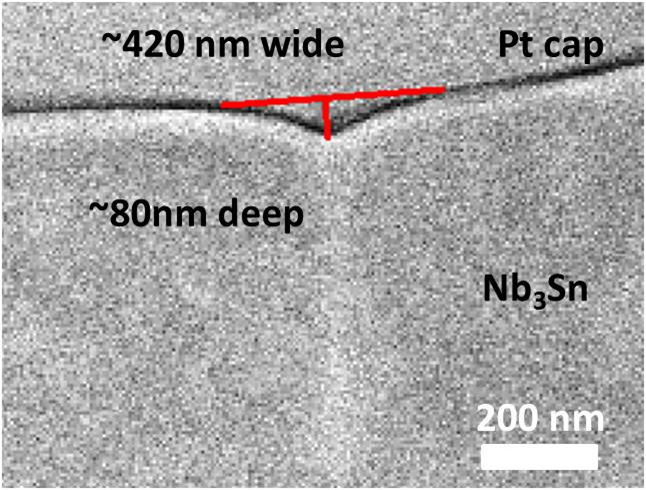}
    \caption{HAADF-STEM image of the cross-section of the top surface of a Nb$_3$Sn GB. }\label{fig:GBdip}
\end{figure}

Finally, we observe divots in the Nb$_3$Sn surface at GBs; the HAADF-STEM image in Fig. \ref{fig:GBdip} displays the geometry of a GB on the top surface. It has $\sim$80 nm of depth and $\sim$420 nm of width. The unique compositional and geometrical changes at GBs may provide pathways for flux to penetrate the surface. This observations motivates us to consider surface geometry as an additional factor in our Ginzburg-Landau modeling of vortex penetration.

\section{\label{sec:TcCalcs}The effect of tin-rich stoichiometry on $T_c$}

\begin{figure}
   \centering
   \includegraphics*[width=.9\columnwidth]{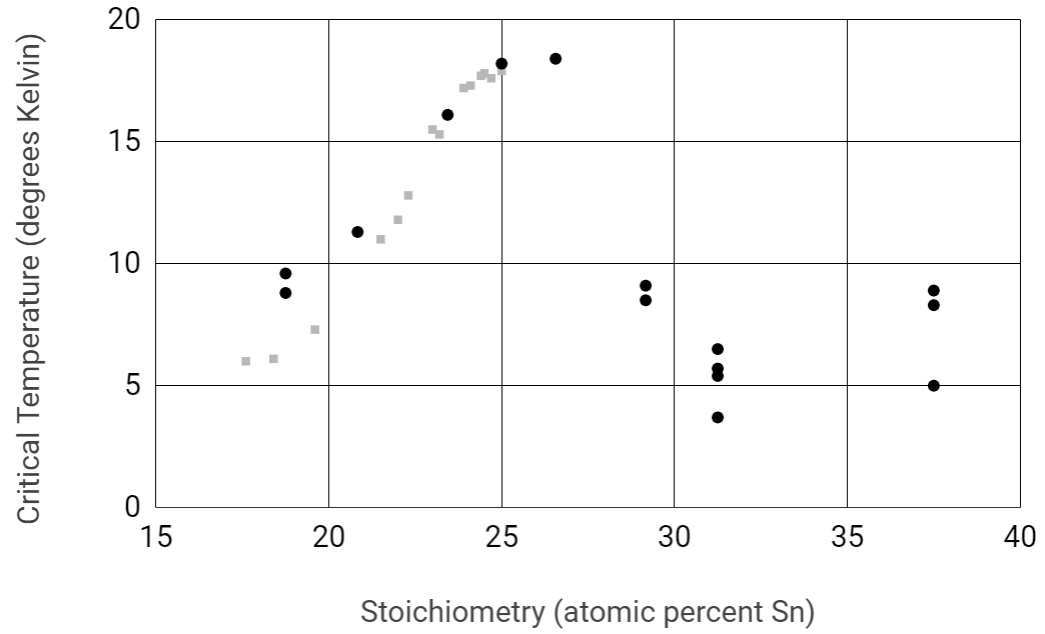}
   \caption{Experimental $T_c$ \cite{Devantay} (grey squares) and calculated $T_c$ (black circles) for A15 Nb-Sn of different stoichiometries. For some stoichiometries, multiple possible defect configurations were considered. The calculated $T_c$ reaches a minimum of about 5 Kelvin in the tin-rich regime.}
   \label{fig:Tc}
\end{figure}

Previous works have demonstrated that both defects and strain can reduce the $T_c$ of Nb$_3$Sn, and that this effect can be attributed to a reduction of the Fermi-level density of states \cite{sitaraman2019,godeke2018fundamental}. At GBs, both of these effects may play an important role, but because these regions are so small, it is difficult to directly probe their superconducting properties. To determine the effect of Sn-antisite defects, we consider \emph{ab initio} $T_c$ values calculated using Eliashberg theory \cite{McMillan} and Density Functional Theory (DFT) \cite{dft,sitaraman2019}. We present such results obtained using a Wannier-based k-point sampling approach\cite{mlwf}. For the experimentally measured stoichiometry range of the A15 phase, the predicted $T_c$ values are similar to or modestly higher than experimental values, as described in Table 1. For experimentally inaccessible tin-rich stoichiometry, $T_c$ values fall to a minimum of about 5K at 31.25\% Sn stoichiometry (Fig. \ref{fig:Tc}). This is well within the range that has been observed around GBs. As expected, we observe a close correlation between calculated $T_c$ and Fermi-level density of states for all calculations.

Our atomistic GB calculations show that the presence of tin antisite defects dramatically widens the reduction in the Fermi-level density of states around the GB, from the range of $\sim$1-1.5 nm described in section~\ref{sec:Experimental} to a range of at least $\sim$2~nm \cite{kelley2020abinitio}. By considering the average Fermi-level density-of-states over the volume of a Cooper pair, we estimate the local suppression of $T_c$ from both a clean GB and a GB containing the observed tin-concentration profiles \cite{lee2019grain}. Our estimates show clean GBs suppress $T_c$ by only about 20\%, but GBs with representative Sn-antisite concentrations suppress $T_c$ by close to 70\% \cite{kelley2020abinitio}. Based on this result, we conclude that the combined impact from anti-site defects and strain is significantly larger than the impact from strain alone. This predicted sensitivity of GB properties to defect segregation further motivates our Ginzburg-Landau modeling of GBs with different local superconducting properties.

\begin{table}[h!t]
    \centering
    \setlength\tabcolsep{3.8pt}
    \caption{Calculated vs. Measured $T_c$}
    \begin{tabular}{@{}llcc@{}}
        \toprule
        \textbf{Composition} & \textbf{Experimental T$_c$ (K)} & \textbf{Calculated T$_c$ (K)} \\
         \midrule
          18.75\% Sn & 6 & 9.2$^\dagger$ \\
         \midrule
          20.83\% Sn & 9.5 & 11.3  \\
         \midrule
          23.44\% Sn & 16 & 16.1  \\
         \midrule
      25.00\% Sn & 18 & 18.2 \\
      \midrule
      31.25\% Sn & n/a & 5.3$^\dagger$ \\
        \bottomrule
    \end{tabular}
    $\dagger$ Averaged over multiple configurations.
\end{table}

\section{\label{sec:TDGL}Time-Dependent Ginzburg-Landau Simulations of vortex nucleation}

\subsection{\label{sec:TDGL_intro}Introduction to Methods}


Time-dependent Ginzburg-Landau (TDGL) theory is sophisticated enough to capture vortex dynamics without becoming too algebraically complicated and computationally expensive.
We solve the TDGL equations using a finite element method implemented in FEniCS\cite{AlnaesBlechta2015a,logg2012automated}.
We follow the finite element formulation described by Gao et.~al\cite{gao2015efficient, pack2019role}.
The equations are
\begin{widetext}
\begin{align}
  \frac{\partial \psi}{\partial t} + i \phi\psi = & -\alpha' \psi + |\psi|^2\psi+\left(\frac{-i}{\kappa_0}\nabla- \mathbf{A}\right)^2\psi\label{eq:TDGL1}\\
  \mathbf{j}= & \nabla\times{\nabla\times{\mathbf{A}}}\nonumber\\
                                                  = &- \frac{1}{u_0}\left(\frac{\partial \mathbf{A}}{\partial t} + \frac{1}{\kappa_0}\nabla\phi\right)-\frac{i}{2\kappa_0}\left( \psi^* \nabla\psi - \psi\nabla\psi^*\right) - |\psi|^2\mathbf{A}\label{eq:TDGL2}\\
    \left( \frac{i}{\kappa_0}\nabla\psi + \mathbf{A}\psi \right)\cdot n = & 0 \text{ on surface}\label{eq:TDGL3}\\
    \left(\nabla\times{\mathbf{A}}\right)\times n = & \mathbf{H}\times n \text{ on surface}\label{eq:TDGL4}\\
    -\left( \nabla\phi + \frac{\partial \mathbf{A}}{\partial t} \right)\cdot n = & 0 \text{ on surface}\label{eq:TDGL5}.
\end{align}
\end{widetext}
These equations depend on the order parameter $\psi$, the magnetic vector potential $\mathbf{A}$, and the electric potential $\phi$, all of which can vary in space and time.
There are also three dimensionless material parameters that are modified from the usual definition order to accommodate material inhomogeneities.
We follow the procedure in reference\cite{pack2019role} and define material parameters with respect to reference point, which, in this case, we take to be the bulk Nb$_3$Sn.
The constant $\kappa_0$ is the Ginzburg-Landau parameter, the ratio of the penetration depth $\lambda_0$ and the coherence length $\xi_0$ in the bulk.
Similarly, $u_0$ is the ratio of the characteristic time scales for variations in the order parameter and current in the bulk.
Finally, Sn segregation is modeled through a spatially varying critical temperature quantified in the parameter $\alpha^\prime = (1 - T/T_c)/(1 - T/T_{c}^{bulk}) \propto 1 - T/T_c$.

The Ginzburg-Landau model has some inherent limitations.
Importantly, it is quantitatively valid only near $T_c$, and the dynamics Eqs.~(\ref{eq:TDGL1})-~(\ref{eq:TDGL2}) are appropriate for gapless superconductors.
The model captures the relevant physics for describing vortex nucleation and pinning while accommodating inhomogenous material parameters, making it a useful model at level of detail for this study.
Here, we use the model as a tool for probing the stability of superconducting configurations in the presence of segregated grain boundaries and varying applied magnetic fields.
The limitations of the model mean that the results of these simulations are not quantitativleiy accurate; however, they do give a qualitative insights into a potential explanation for the Q-slope phenomenon.
A model for gapped superconductors introduces a second time scale\cite{kramer1978theory}, and future work may incorporate these more realistic dynamics.

This formulation reduces the full three-dimensional problem into a two-dimensional problem by assuming symmetry along the z-axis. The magnetic field points along the z-axis, fixing variations in the order parameter and magnetic vector potential to the x-y plane. 

\begin{figure}
    \includegraphics[width=.9\columnwidth]{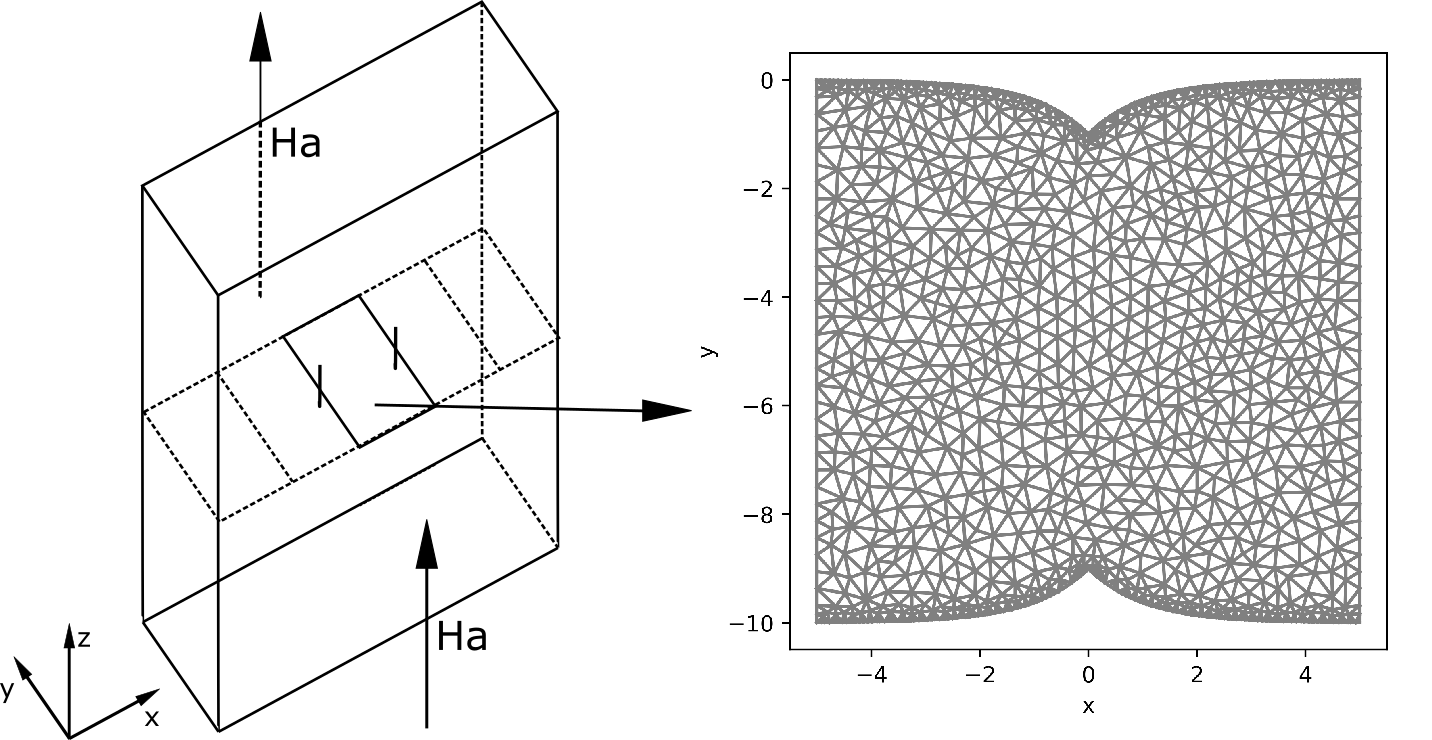}
    \caption{ The black square interior to the film geometry marks our domain of simulation. It lies perpendicular to the applied magnetic field $H_a$. We assume there are no variations in the direction of $H_a$ so that we can simulate a 2D domain.}\label{fig:filmGeometry}
\end{figure}

For all simulations in this section, we consider the film geometry seen in Figure \ref{fig:filmGeometry}. We take a rectangular cross-section lying in the x-y plane. We enforce periodic boundary conditions on the left and right side. On the top and bottom, we enforce Dirichlet boundary conditions for the magnetic field and Neumann boundary conditions for the order parameter. 

We model geometric defects by removing an exponential cut out from the top and bottom of the film. The region removed is given by $d e^{-\frac{|x|}{w}}$ where $d$ is the depth of the cut out and $w$ determines the width. The depth and width are chosen to match experimentally observed geometries. 

\begin{figure}
    \includegraphics[width=\linewidth]{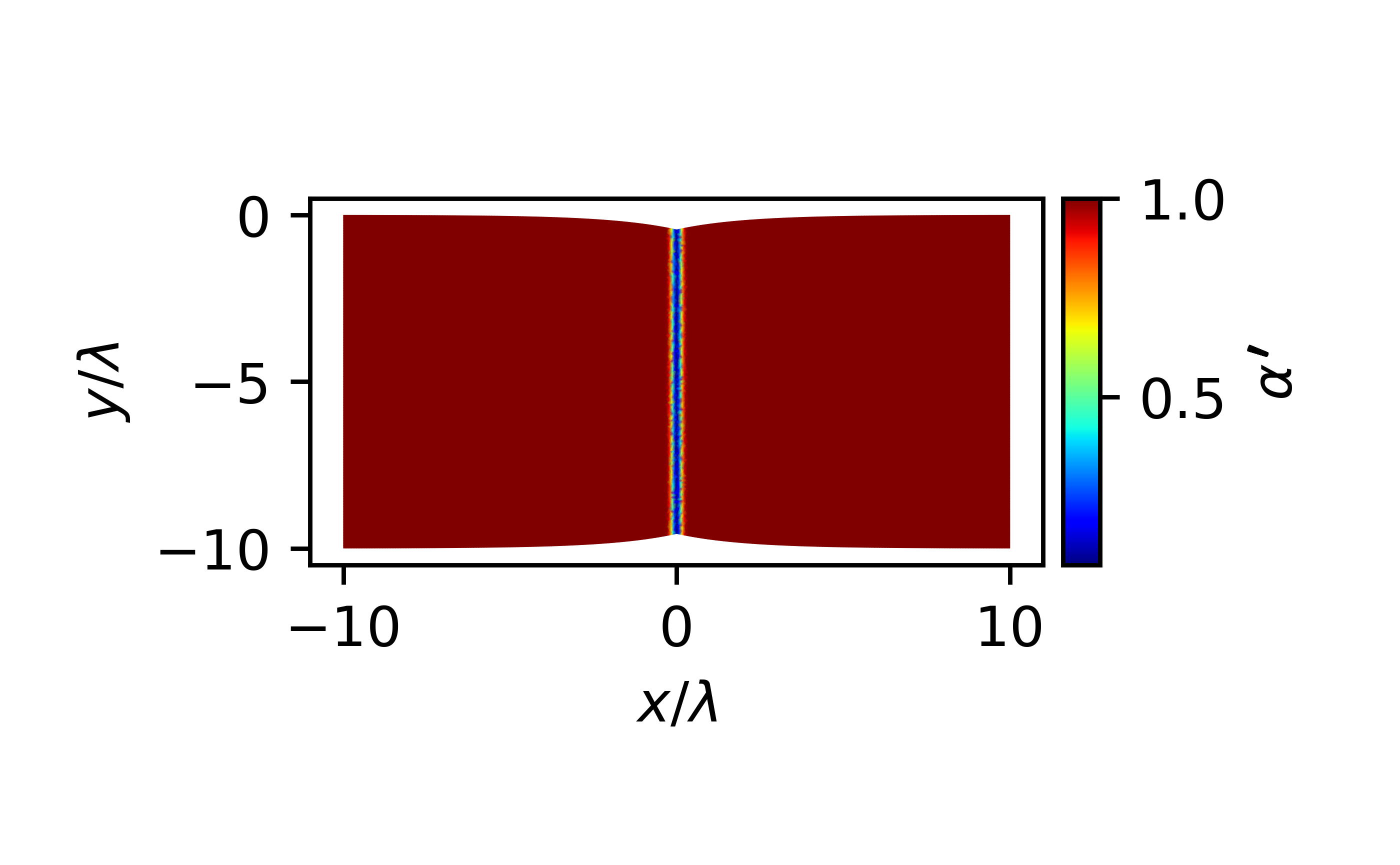}
    \caption{Projection of $\alpha' = 1 - T/T_c$ onto a mesh, where $\alpha'=0$ (i.e., $T \approx T_c$) in the region $|x|/\lambda \leq 0.125$ and $\alpha'=1$ elsewhere. The width of this region comes from experimental observations.}\label{fig:alphaProfile}
\end{figure}

To capture Sn segregation we allow $T_c$ to vary over the domain through the parameter $\alpha^\prime$.
We take $T_c^{bulk}$ = 19 K and $T$ = 4.4 K, so that $T/T_c = 1 - 0.77 \alpha'$.
To mimic the distribution of material inhomogeneities shown in Figure \ref{fig:GBsegregation}, we let $\alpha^\prime \leq 1$ in the GB region $|x| \leq \xi/2$ and $\alpha^\prime=1$ elsewhere.
The projection of this onto the mesh is shown in Figure \ref{fig:alphaProfile}. 

In addition to stoichiometry, previous work suggests that strain effects in the neighborhood of a grain boundary can also effect local superconducting properties.
For example, a 1\% strain can result in a 10\% reduction in the Fermi level density of states\cite{godeke2018fundamental}.
In contrast, an excess of 5\% Sn can result is more than a 50\% reduction in the density of states at the Fermi level.
Here, we focus on the dominant, stoichiometry effect and model a reduction in $T_c$ that is strongly localized to the grain boundary.

The Ginzburg-Landau parameter for Nb$_3$Sn is $\kappa = \lambda / \xi \sim$26, which is challenging to simulate because of the extreme separation in length scales that require a very refined mesh.
However, the relevant physics for vortex nucleation are variations in material parameters on length scales comparable to the superconducting coherence length, $\xi$.
Therefore, we have simulated a moderate type-II material ($\kappa = 4$) but scaled the width of the segregated region (i.e., depleted $T_c$) so that its dimensions relative to $\xi$ are the same as that observed in Figure~\ref{fig:GBsegregation}.
At sufficiently high fields, the line of vortices undergoes a buckling instability when the vortex-vortex repulsion overcomes the pinning strength of the grain boundary.
Because our simulations use a smaller value of the penetration depth, we under-estimate the vortex-vortex repulsion and overestimate the critical field at which the chain becomes unstable.
However, the pinning force in real-world cavities also depends on details, such as the slope of the $\alpha'$ in the transition region, as we discuss below.
These details will affect quantitative predictions for real-world cavities; however, we expect these simulations to capture the qualitative picture.


\subsection{\label{sec:TDGL_BoundaryNucleation}Vortex Nucleation in Grain Boundaries}

To simulate the nucleation of vortices into a grain boundary, we set the value of the magnetic field at the top and bottom boundary such that it is low enough that an array of vortices do not penetrate directly into the bulk, but large enough for vortices to enter into the grain boundary\cite{pack2019role}.
As we evolve in time (assuming a constant applied field), two different behaviors are observed depending on the magnitude of the applied field.
In the first scenario, vortices enter into the grain boundary at the geometric divot.
With increasing field, the spacing between vortices decreases until it reaches critical levels.
In the second scenario, vortices first fill the grain boundary, as in the first case, but then begin to penetrate into the grain from the boundary.

\begin{figure}
\centering
    \includegraphics[trim={0, 25, 0, 25}, clip, width=\columnwidth]{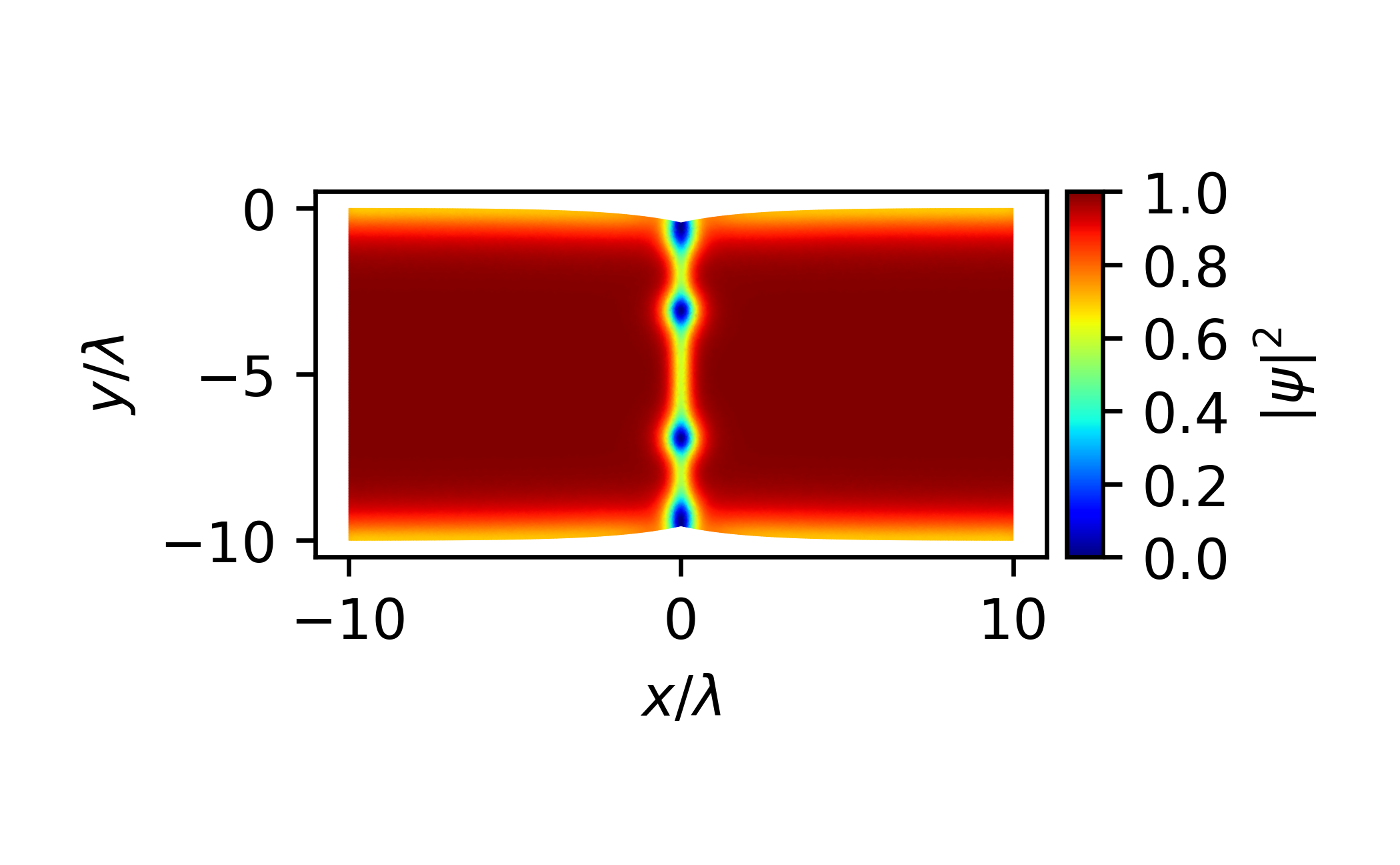}
    \includegraphics[trim={0, 25, 0, 25}, clip, width=\columnwidth]{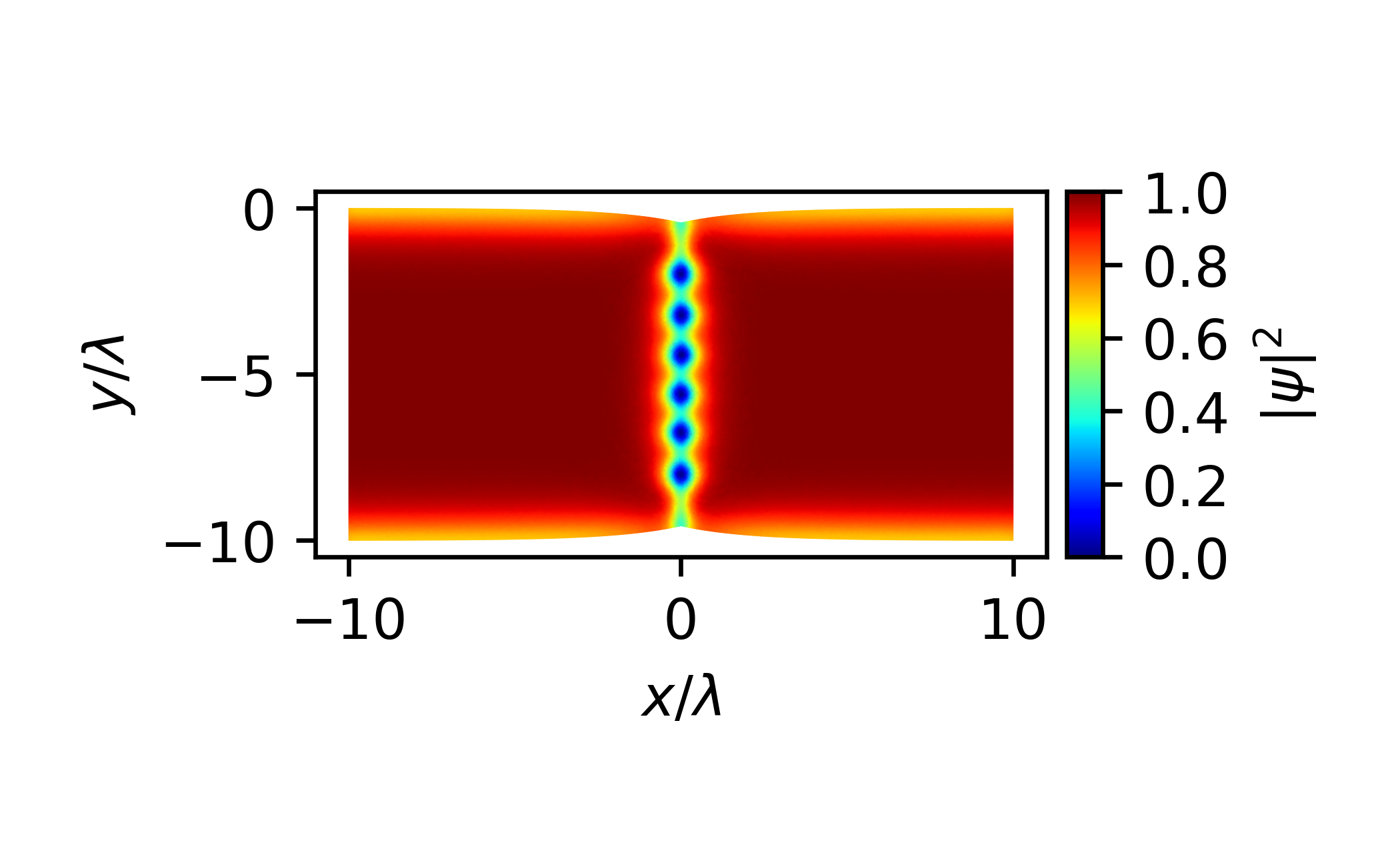}
    \includegraphics[trim={0, 25, 0, 25}, clip, width=\columnwidth]{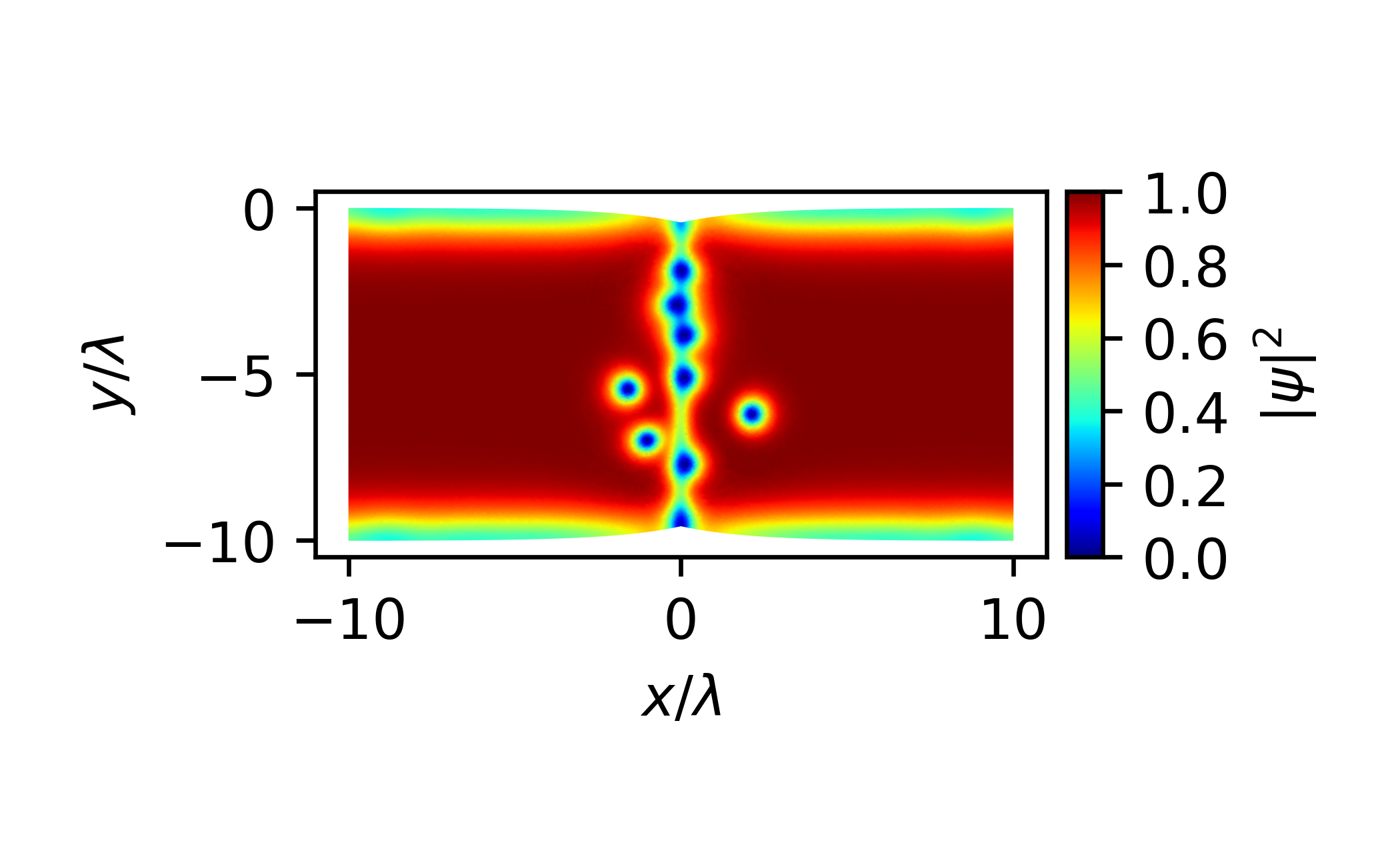}
    \includegraphics[trim={0, 25, 0, 25}, clip, width=\columnwidth]{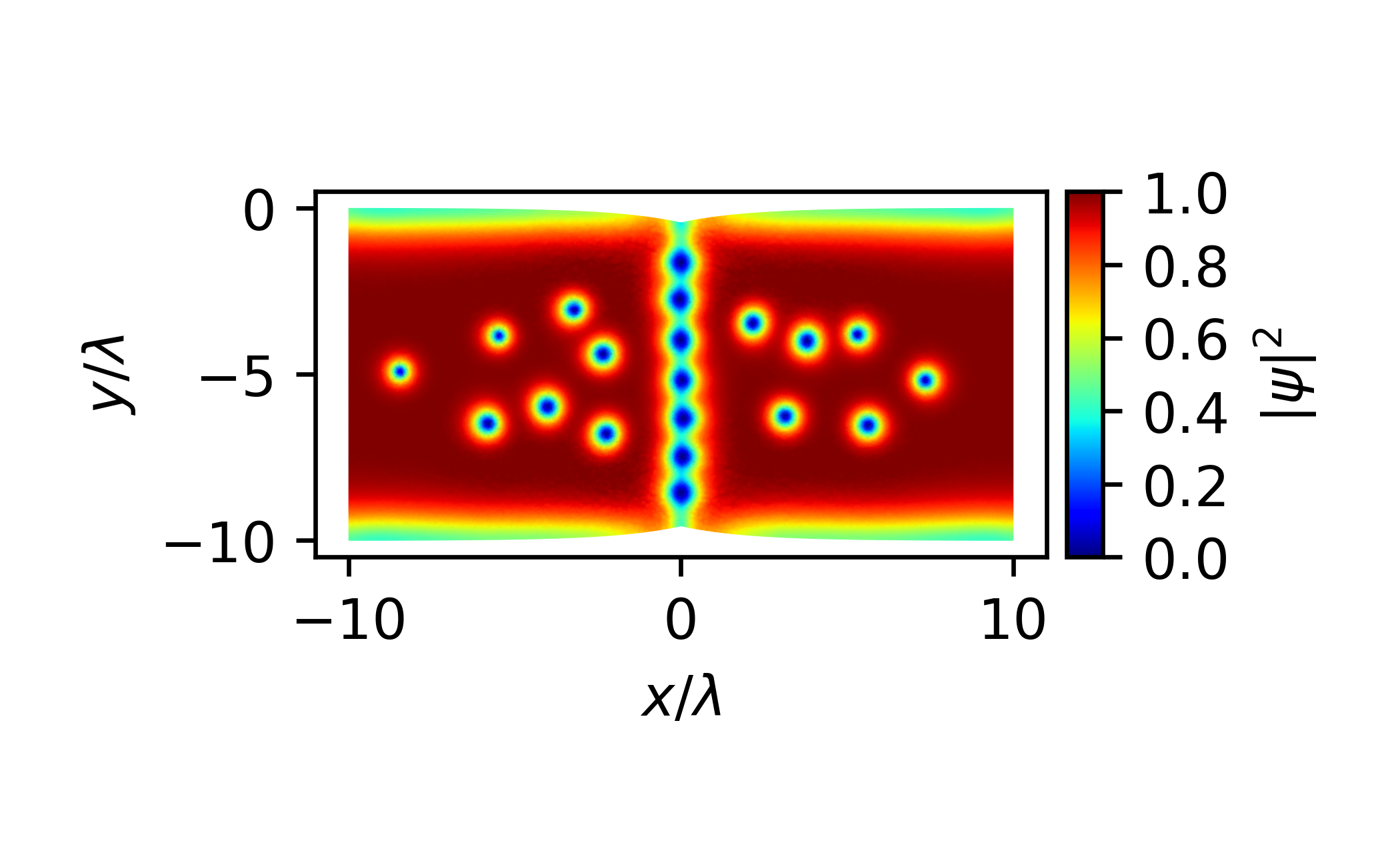}
    \caption{Sequence of behaviors for increasingly large applied magnetic fields.
      An applied magnetic field is set as a boundary condition to the top and bottom of the film, with periodic boundary conditions on the right and left sides.
      At moderate applied magnetic fields, vortices penetrate only into the grain boundary from the geometric defect (top).
      At higher fields, a critical vortex spacing is reached (second pane) above which vortices begin to penetrate the bulk from the grain boundary (third pane).
      Finally, having entered the bulk, vortices disperse and fill the grain with an equilibrium distribution (bottom).}\label{fig:GrainBoundaryPen}
\end{figure}

Once a vortex has penetrated into the grain boundary, it is pushed from the surface, allowing more vortices to come in after it.
Once space is available, another vortex penetrates.
This continues until the vortices have filled the grain boundary, i.e. an optimal spacing between the vortices inside grain boundary has been achieved.
This is illustrated by the sequence in Figure \ref{fig:GrainBoundaryPen}.
Note that vortices are manifest as regions in which the order parameter is reduced to near zero at their center and exponentially decay radially outward to unity.

\begin{figure}
    \includegraphics{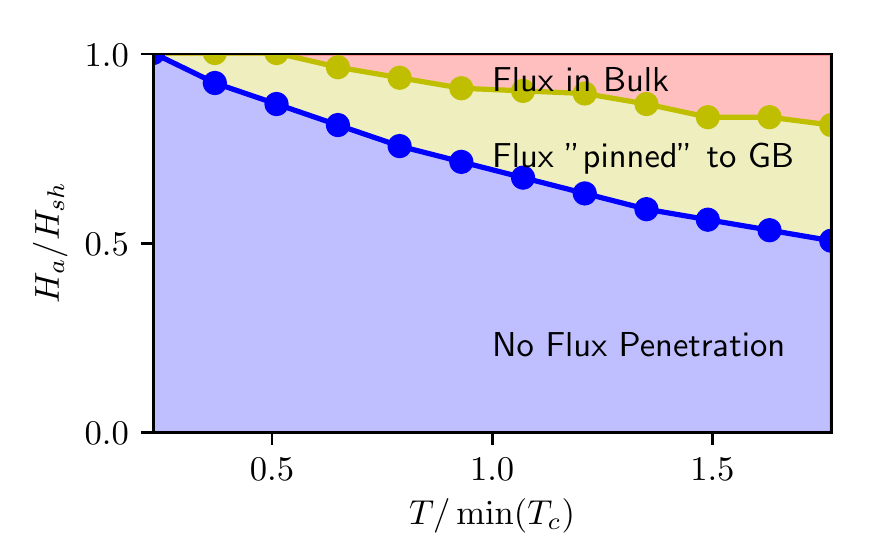}
    \caption{Phase diagram of TDGL predictions for flux penetration in the presence of the grain boundary as a function of the the applied field and minimum critical temperature inside the grain boundary.
      At small applied fields, no flux penetrates (blue).
      At intermediate fields, flux penetrates but is constrained to the grain boundary (yellow).
      At sufficiently high fields, the flux penetrates from the grain boundary in to the bulk material (red).}\label{fig:GLTPhaseDiagram}
\end{figure}

If the applied magnetic field is sufficiently high, vortices will also begin to penetrate into the bulk once the grain boundary has been filled.
See, for example, the bottom of Figure \ref{fig:GrainBoundaryPen}.
The escape of vortices into the bulk occurs through a buckling transition similar to that observed in anisotropic, layered superconductors\cite{dodgson2002phase}.
The field at which vortices penetrate from the grain boundary into the grain is dependent on the distribution of $\alpha' = 1 - T/T_c$. The shallower the slope of $\alpha'$ the lower the applied field needs to be to nucleate vortices into the grain from the grain boundary.
These results are summarized in the phase diagram in Figure~\ref{fig:GLTPhaseDiagram}.
Comparing with results from section~\ref{sec:TcCalcs}, for a segregated region with $T_c \sim 5K$ in a cavity operating at $T = 4.2K$ ($T/T_c \sim 1$), we observe a non-trivial region of the phase diagram that admits flux trapped at the grain boundary.

\begin{figure}
   \includegraphics{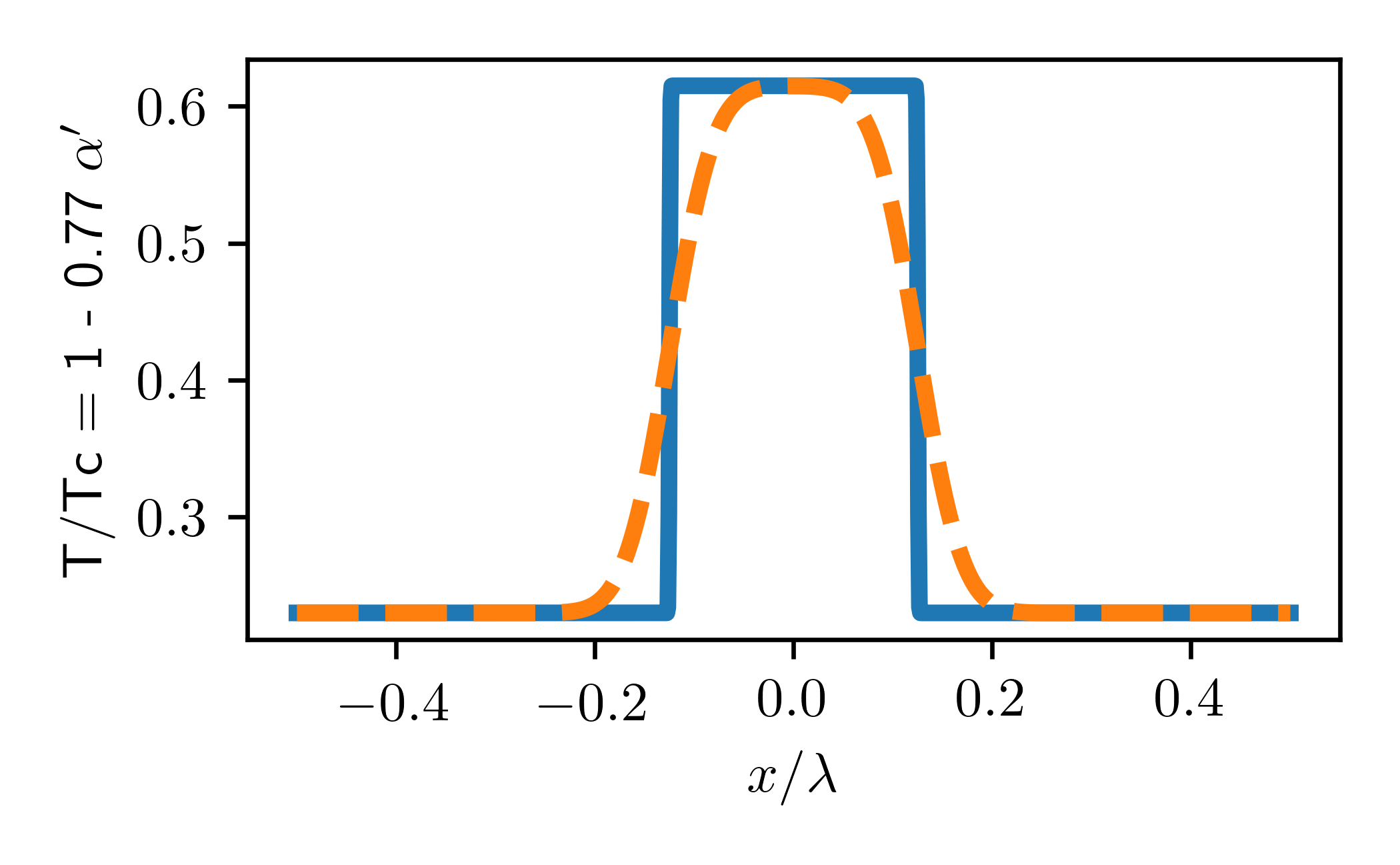}
  \caption{Profiles of potential $\alpha^\prime$ for the transition between the segregated and non-segregated regions.
    Sharp transitions (such as the blue solid curve) keep the vortices constrained to the boundary for larger applied fields.
    A more shallow transition (such as the orange dashed curve), however, allow vortices to escape into the grain more easily.}\label{fig:alphaprofile1D}
\end{figure}

The value of the applied field at which the vortices first leave the grain boundary and penetrate the bulk depends on the properties in the transition zone between the segregated and non-segregated region.  If the transition form $\alpha^\prime < 1$ (segregated region) to $\alpha^\prime = 1$ (non-segregated region) is very sharp (as the blue solid curve in Figure \ref{fig:alphaprofile1D}), then vortices will be trapped in the grain boundary for larger fields.  However, if the transition is more gradual (such as the orange dashed curve), then it is easier for vortices to escape the boundary and penetrate the bulk.  Figures~[\ref{fig:alphaProfile}, \ref{fig:GrainBoundaryPen}, \ref{fig:GLTPhaseDiagram}] were generated with a very sharp interface.

\section{Estimates of vortex dissipation at grain boundaries}
	\label{sec:VortexTheory}

\begin{figure}[!ht]
		\centering
		\includegraphics[width=\linewidth]{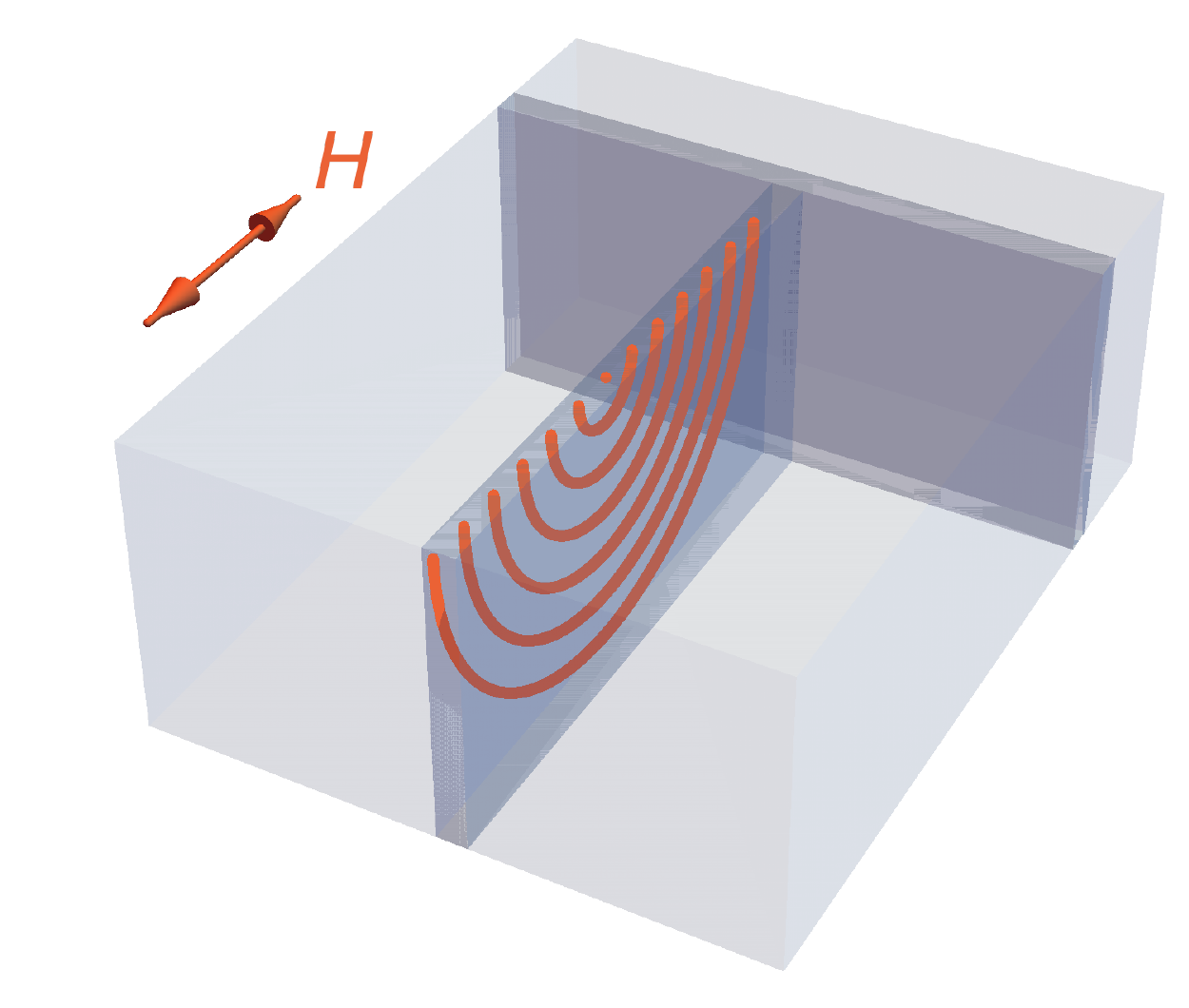}
		\caption{Illustrating vortex nucleation (red lines) in a superconductor (light
		gray region) subject to a surface magnetic field $H$.
		Vortex entry starts at regions where superconductivity is weakest (dark gray
		regions, here representing grain boundaries).}
		\label{fig:GBNuc}
	\end{figure}

	Inhomogeneous properties of superconductors have high impact on the performance of
	SRF cavities, affecting figures of merit such as the residual resistance due to trapped
	magnetic flux~\cite{GurevichCio2013,HallSet2018,LiarteSet2018}, and the superheating
	field~\cite{Kramer1968,catelani2008temperature,transtrum2011superheating,LiarteSet2016,LiarteSet2017,
	NgampruetikornSau2019}.
	Grain boundaries well aligned with the surface magnetic field can become weak spots for
	the nucleation of superconducting vortices (see Fig.~\ref{fig:GBNuc}), and could be
	ultimately responsible for the quench of an SRF cavity.
        Note that our Ginzburg-Landau model did not simulate the full semi-loops of Figure~\ref{fig:GBNuc}.
	In this section, we discuss simple estimates for the power dissipation and heat diffusion
	due to nucleation of vortices at grain boundaries in Nb$_3$Sn cavities.

	We start with an estimate for the power dissipated in an SRF cavity when a grain
	boundary is filled with $n$ superconducting vortices.
	We assume of order $\mathcal{O}(n)$ vortex lines are annihilated once per cycle,
	their energy is lost, and the power dissipated per vortex line is simply the drop in
	the energy of the outside magnetic field times the RF frequency.
	Our estimate gives a rough estimate for the actual power
	dissipated by vortices at grain boundaries, if the field reaches
	high enough values for them to enter.
	 
	The drop in magnetic energy when a vortex line of length $D$ nucleates into the
	superconductor is given by:
	\begin{align}
		\Delta E
			& = \left| \int \frac{1}{2 \mu}\left(B -\frac{\Phi_0 D}{V}\right)^2 dV
				- \int \frac{B^2}{2 \mu} dV \right| \nonumber \\
			& \approx \frac{B \Phi_0 D}{\mu},
	\end{align}
	where $V$ is the volume, $\mu$ is the permeability of free space, $\Phi_0$ is the
	fluxoid quantum, and $\lambda$ is the penetration depth.
	Note that $\Delta E$ is also approximately the work done by the external magnetic field to
	push a vortex into the bulk of the superconductor:
	$W \approx f_L \cdot D \cdot \lambda = ( \Phi_0 B / (\mu \lambda) )D \lambda = \Delta E$,
	where $f_L$ is the Lorentz force per length.
	Thus, our calculation gives the vortex dissipation at grain
	boundaries assuming that the vortices do not leave the grain boundary
	as the external field drops and changes sign, as our simulations
	indicate, and as one would expect for vortices that enter a distance
	more than $\lambda$, past the surface nucleation barrier.
	The total energy drop for a grain boundary of linear size $D$ with vortices
	spaced by $\lambda$ (see Fig.~\ref{fig:GBNuc}) is
	$\Delta E_{GB} \approx \Delta E (D / \lambda) = B \Phi_0 D^2 / (\mu \lambda)$, and the
	power dissipated per grain boundary is given by
	\begin{equation}
		P_{GB}
			\equiv f \Delta E_{GB} =  \frac{B \Phi_0}{\mu} \frac{f D^2}{\lambda},
	\end{equation}
	where $f$ is the cavity frequency.
	For a 1.3GHz Nb$_{3}$Sn cavity with typical grain size of $D \approx 1 \mu$m, we
	find $P_{GB} \approx 621$nW at $B = 60$mT.

	Note that our estimate relies on the assumption that vortices quickly fill the grain
	boundary before being annihilated during the RF cycle.
	If the vortex line relaxes slowly, the RF field might vanish and change sign before
	vortices had time to fill the grain boundary, and our assumption would not be valid.
	We now show that that is not the case --- vortices move at extremely high speeds in the
	typical environment of Nb$_3$Sn SRF cavities.

\begin{figure}[!ht]
		\centering
		\includegraphics[width=\linewidth]{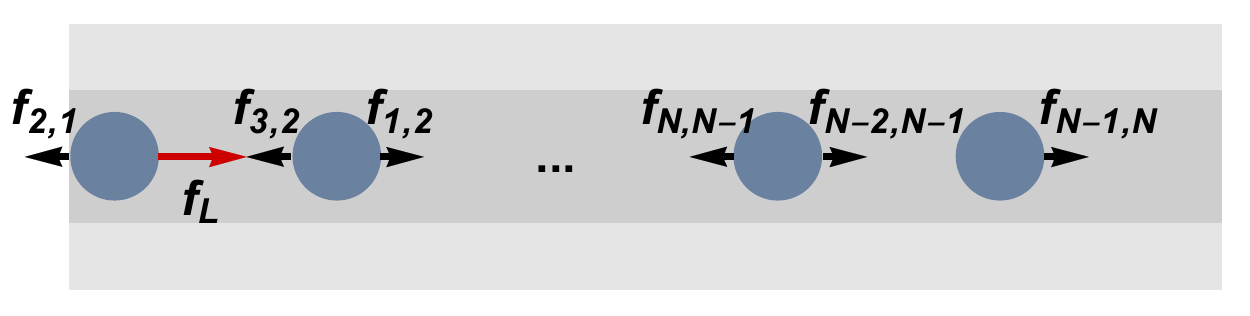}
		\caption{Illustrating a ``train'' of $N$ vortex lines (blue disks) moving through a grain
		boundary (dark gray) of a superconductor (light gray).
		The first vortex is subject to a surface Lorentz force from the RF field and each vortex
		is repelled by its nearest neighbors.}
		\label{fig:VortexTra}
	\end{figure}

Consider the one-dimensional motion of a ``train'' of $N$ vortices moving through a
	grain boundary towards the superconductor interior (see Fig.~\ref{fig:VortexTra}).
	Assuming over-damped dynamics, the equations of motion for each vortex line $i$ of
	velocity $v_i$ read:
	\begin{align}
		& \eta_0 v_1
			= f_L - f_{2,1}, \nonumber \\
		& \eta_0 v_2
			= f_{1,2} - f_{3,2}, \nonumber \\
		& \quad \vdots \nonumber \\
		& \eta_0 v_{N-1}
			= f_{N-2,N-1} - f_{N,N-1}, \nonumber \\
		& \eta_0 v_N
			= f_{N-1,N}, \nonumber
	\end{align}
	where $f_L = \Phi_0 H_{rf} / \lambda$ is the Lorentz force per length at the surface,
	$H_{rf}$ is the surface magnetic field, $\eta_0 = {\Phi_o}^2/(2\pi {\xi}^2 \rho_n)$ is the
	Bardeen-Stephen viscosity~\cite{BardeenSte1965}, $\rho_n$ is the resistivity of the
	normal state and $f_{i,j}$ is the repulsion force from vortex $i$ into vortex $j$.
	Thus,
	\begin{equation}
		v_{BS}
			\equiv \frac{1}{N} \sum_{i=1}^N v_i
			= \frac{f_L}{N \eta_0}
			= \frac{2 \pi}{\mu \Phi_0} \frac{\rho_n \xi^2}{\lambda} \frac{B_{rf}}{N}.
	\end{equation}
	For Nb$_3$Sn at $B_{rf} = 60$mT, we find $v_{BS}= 2.4\mu$m/ns and
	$v_{BS}= 24\mu$m/ns for ten vortices and one vortex, respectively.
	The average velocity of the vortex train is high enough for vortices to quickly fill in the grain
	boundary during the RF cycle, but this numerical value should be taken with a grain of salt.
	The Bardeen-Stephen formula is not valid at these high speeds, which also exceed the
	pair-breaking limit of the superconducting condensate~\cite{EmbonZel2017}.
	In principle, one could incorporate pair-breaking mechanisms into our simple
	estimates by using the nonlinear viscosity calculated by Larkin and
	Ovchinnikov (LO)~\cite{LarkinOvc1975}: $\eta (v) = \eta_0 / [1+(v/v_0)^2]$, where $\eta_0$ is
	the Bardeen-Stephen viscosity and the LO velocity $v_0$ marks the onset of highly dissipative
	states for straight vortices.
	Following the same steps that we have used to calculate $v_{BS}$, and assuming that each
	vortex moves with the same vortex-train velocity $v_{LO}$, we find
	\begin{align}
		v_{LO}
			& = \frac{{v_0}^2}{2 \, v_{BS}} \left(1-\sqrt{1-4\frac{{v_{BS}}^2}{{v_{0}}^2}} \right)
				\nonumber \\
			& \approx v_{BS} \left(1 + \frac{{v_{BS}}^2}{{v_0}^2} \right),
		\label{eq:vlo}
	\end{align}
	for $v_{BS} \ll v_0$, showing that nonlinear corrections lead to an increase of the velocity of
	the vortex train from the Bardeen-Stephen result.
	However, note that estimates for Nb$_3$Sn at low temperatures~\cite{GurevichCio2008} yield
	$v_0 \approx 0.1$km/s, which is much smaller than our estimates for $v_{BS}$, so that a more
	sophisticated analysis is needed to better describe vortex motion at these high-speed regimes.
	In recent work~\cite{PathiranaGur2020}, Pathirana and Gurevich consider the nonlinear
	dynamics of curvilinear vortices subject to LO viscous forces to study dissipation and vortex
	teardown instabilities.

	Grain boundary activation might be associated with the degradation of the quality factor $Q$
	of SRF cavities at high fields.
	We now use our estimates to calculate the number of active grain boundaries needed to
	deplete $Q$ by a certain amount.

	The quality factor is given by $Q=G {B_{rf}}^2 / (2 \mu^2 P)$, where $P$ is the dissipated
	power per unit area and $G$ is a geometry factor.
	We break up the total surface area $s$ of the cavity into $N$ blocks, so that $s = N s_{GB}$,
	where $s_{GB}$ is the average area occupied by one grain boundary.
	Assume inactive and active blocks dissipate power $P_1$ and $P_1 + P_{GB}$, respectively,
	where by active block we mean a block with a grain boundary filled with vortices.
	For $M$ active blocks,
	\begin{align}
		Q
			& = \frac{G {B_{rf}}^2}{2 \mu^2 (NP_1 +MP_{GB})/ (Ns_{GB})} \nonumber \\
			& = \frac{G {B_{rf}}^2 s_{GB}}{2 \mu^2 (P_1 +xP_{GB})},
		\label{eq:Qeq1}
	\end{align}
	where $x\equiv M/N$ is the ratio of active grain boundaries.
	In the absence of active grain boundaries, we assume $Q = Q_1$ is constant
	(i.e.~$P_1 \sim {B_{rf}}^2$), so that
	\begin{equation}
		P_1
			= \frac{G {B_{rf}}^2 s_{GB}}{2 \mu^2 Q_1}.
		\label{eq:P1eq}
	\end{equation}
	Plugging Eq.~\eqref{eq:P1eq} into Eq.~\eqref{eq:Qeq1} and solving for $x$, we find
	\begin{equation}
		x
			= \frac{G s_{GB} {B_{rf}}^2}{2 \mu^2 P_{GB}}
				\left(\frac{1}{Q}-\frac{1}{Q_1}\right).
	\end{equation}

\begin{figure}[!ht]
		\centering
		\includegraphics[width=\linewidth]{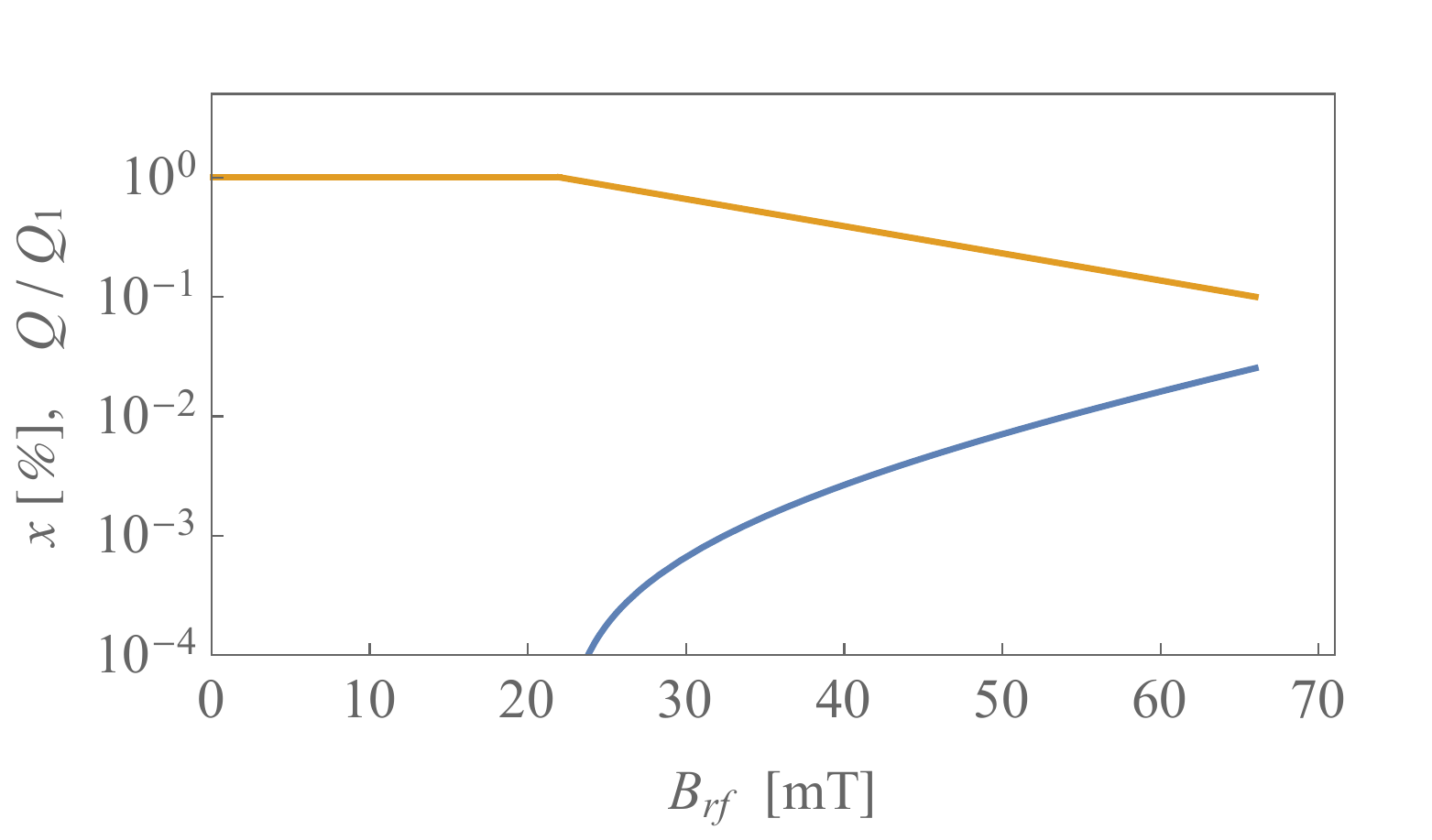}
		\caption{Estimated percentage of active surface grain boundaries (blue curve) as a
		function of $B_{rf}$, corresponding to the quality factor profile displayed in the yellow
		curve for Nb$_3$Sn with $Q_1 = 10^{10}$.}
		\label{fig:QxVsB}
	\end{figure}

	To estimate the percentage of active grain boundaries that is needed for $Q$ to drop by a
	certain amount, we assume Q is constant and equal to $Q_1=10^{10}$ for fields up to
	$B_{rf}=22$mT, and then decreases exponentially until it reaches $Q=10^9$ at
	$B_{rf}=66$mT.
	This form for the $Q$-slope profile is shown in yellow in Fig.~\ref{fig:QxVsB}, and is chosen
	to roughly mimic the experimental data shown with red symbols in Figure~\ref{fig:QvsE}.
	Here we assume that $Q$ degradation is solely due to grain activation; the field of vortex
	penetration is chosen to roughly match that of the onset of the experimental Q slope
	($\approx 22$mT).
	Note that Calculations for the exactly solvable limit of Josephson vortices which penetrate
	along weakly-coupled planar junctions yield a threshold critical current density
	$J_c \sim \phi_0/(\lambda \,{\lambda_J}^2)$, where $\lambda_J$ is the Josephson penetration
	depth, which is assumed to be much larger than the penetration
	depth~\cite{SheikhzadaGur2017}.
	Figure~\ref{fig:QxVsB} shows a plot of the percentage of active grain boundaries ($100\, x$,
	blue curve) as a function of $B_{rf}$ corresponding to the artificial $Q$-slope profile shown
	in the yellow curve (using Nb$_3$Sn parameters with $Q_1=10^{10}$,
	$s_{GB} = 0.5 \mu$m$^2$ and $G=278 \Omega$).
	Note that about 0.03\% of the surface grain boundaries need be filled with vortices for $Q$
	to drop from $10^{10}$ to $10^9$ for Nb$_3$Sn at about 66mT. 

\begin{figure}[!ht]
		\centering
		\includegraphics[width=\linewidth]{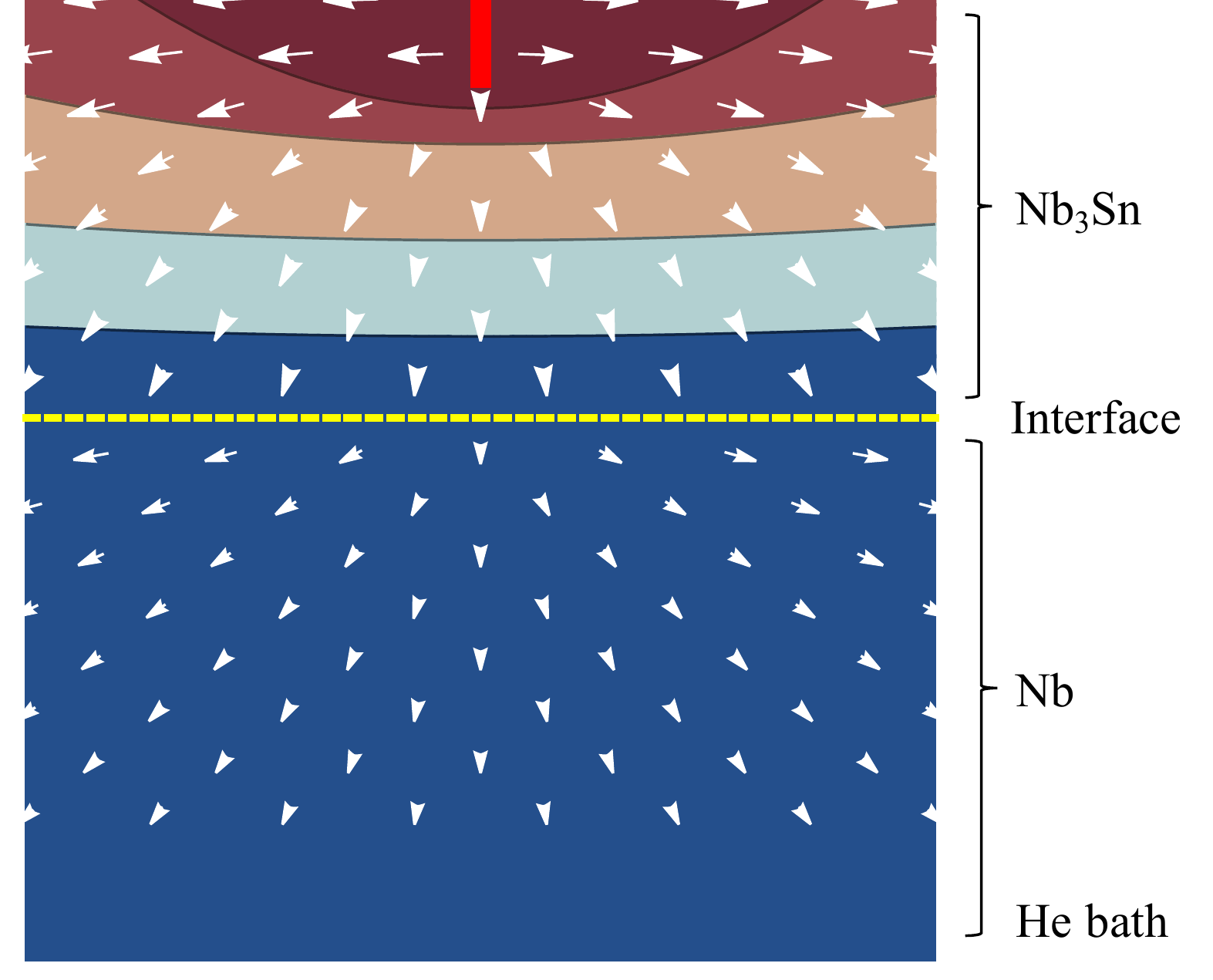}
		\caption{Sketch of our model for heat diffusion near an active grain boundary (red
		rectangle at the top center).
		Red and blue tones represent contour regions of high and low temperatures,
		respectively.
		The dashed yellow line represent the interface between Nb, whose outer surface is in
		contact with the He bath (bottom of the figure), and Nb$_3$Sn, whose inner surface
		is in contact with vacuum and the rf field (top).
		White arrows represents approximate directions for the heat flux.
		In our model, heat fronts move along one and three dimensions for distances smaller
		and greater than the grain size $D$, respectively.
		The change in the direction of heat flux at the interface is due to the wide separation
		of the thermal conductivities of Nb and Nb$_3$Sn.}
		\label{fig:HeatDif}
	\end{figure}

	We end this section with a simple model calculation of the
	steady-state thermal heating at a grain boundary.
	Figure~\ref{fig:HeatDif} illustrates our model for thermal diffusion near an active grain
	boundary.
	Cross sections of Nb and Nb$_3$Sn layers are illustrated below and above the yellow
	dashed line in Fig.~\ref{fig:HeatDif}, respectively.
	The Nb surface is in contact with a low-temperature He bath.
	The Nb$_3$Sn surface is in contact with vacuum, and is subject to a parallel oscillating
	magnetic field.
	The grain boundary is represented by a red rectangle at the top center of
	Figure~\ref{fig:HeatDif}, and has linear size $D$.
	White arrows depict approximate directions for heat diffusion in our model.
	First, we assume one-dimensional heat diffusion away from the grain boundary up to a
	distance $r\approx D$.
	We expect the heat front to attain a semi-spherical shape for distances $r \gtrapprox D$.
	We then assume three-dimensional heat diffusion away from a half-sphere of radius
	$D$ for distances $D \leq r \leq R_1$.
	For $r\leq R_1$ we consider the Nb$_3$Sn thermal conductivity $\kappa = \kappaNbSb$.
	At last, we assume three-dimensional heat diffusion away from a half-sphere of radius
	$R_1$ for distances $R_1 \leq r \leq R_2$, with $\kappa$ given by the Nb thermal
	conductivity $\kappaNb$.
	Note that our assumptions are stronger when $R_2 \gg R_1$ ($R_2 / R_1 \sim 10^3$
	for typical Nb$_3$Sn/Nb SRF cavities).

	The equilibrium temperature profile can then be cast from the stationary solutions of the
	heat equation in each region:
	\begin{align}
		T(r)
			=
			\begin{cases}
				\zeta (r), & \text{for } 0 \leq r \leq D, \\
				c/r + d, & \text{for } D \leq r \leq R_1, \\
				e/r + g, & \text{for } R_1 \leq r \leq R_2,
			\end{cases}
		\label{eq:TSol}
	\end{align}
	where $c$, $d$, $e$ and $g$ are constants, and $\zeta (r)$ is the stationary solution of the
	one-dimensional heat equation (the Nb$_3$Sn thermal conductivity strongly varies with
	temperature in this region~\cite{CodyCoh1964}, which complicates the problem of finding an
	analytical solution for $\zeta (r)$)~%
	\footnote{Note that the temperature dependence of the thermal conductivity of Nb$_3$Sn
	should also change the stationary solution of the three-dimensional heat equation for
	$D \leq r \leq R_1$.
	Here we ignore this effect for the sake of mathematical simplicity, since our qualitative
	conclusions are not altered if we incorporate the temperature dependence of $\kappa$ in this
	region.}.
	Note that we relax the definition of the coordinate $r$ here, which should be interpreted as
	a lateral distance away from the grain boundary for $0 \leq r \leq D$, and a depth coordinate
	towards the Helium bath for distances $r>D$.
	
	To calculate $\zeta (r)$, we use Fourier's law  --- $\dot{\mathcal{Q}} = - \kappa \, d T / dr$,
	where $\dot{\mathcal{Q}}$ is the heat flux.
	The stationary solution of the heat equation can be found from the solution of
	$ d \dot{\mathcal{Q}} / dr =0$, {\it i.e.}
	\begin{align}
		-\kappaNbSb (T) \, \frac{dT}{dr}
			= a,
		\label{eq:NonlinearFourier}
	\end{align}
	where $a$ is constant.

        The thermal conductivity $\kappaNbSb(T)$ in the superconducting layer 
	has two important contributions: a phonon contribution and an electronic component (carried by superconducting quasiparticles). 
	The low-temperature phonon thermal conductivity is strongly
	dependent on the morphology of the crystal~\cite{CahillPoh1988};
	in clean insulating crystals it is dominated by scattering
	off grain boundaries and sample boundaries, and varies as 
	$T^3$. Scattering off impurities can cut off the contribution
	of high-frequency phonons, or even resonantly cut off certain
	frequency bands. All of these mechanisms lead to a thermal
	conductivity that monotonically increases with temperature, so
	we avoid the complexity by using a constant phonon thermal
	conductivity $k_1$, giving a lower bound for the conductivity
	and hence an upper limit to the heating. The electronic portion
	of the thermal conductivity $k_2$ in the normal metal at low 
	temperatures is roughly independent of temperature, and is
	set by the electronic mean-free path. In the superconductor, 
	it decreases exponentially as $\exp(-\Delta (0)/k_B T)$, 
	as seen experimentally~\cite{CodyCoh1964}.
	Using the BCS relation between the gap and the transition temperature,
	we therefore use
	\begin{equation}
		\kappaNbSb (T)
			= k_1 + k_2 \exp{(-1.76 \, T_c/T)}.
		\label{eq:ThermalConductivity}
	\end{equation}
	We use the normal electron thermal conductivity 
	$k_2 = 2 \times e^{1.76}$ W/m$\cdot$K from~\cite{CodyCoh1964}.
	Because the electronic contribution is negligible at the operating
	temperature of the cavity, we set $k_1=10^{-2}$W/m$\cdot$K as the
	approximate total thermal conductivity of Nb$_3$Sn at
	2K~\cite{CodyCoh1964}.
	Both of these constants are dependent upon the preparation of the
	film, and also could vary from one region of the film to another
	as the growth conditions or the underlying Nb grain orientations
	vary.

	Integration of Eq.~\eqref{eq:NonlinearFourier} results in
	\begin{equation}
		\Pi \, (T)
			= -a\, r +b,
		\label{eq:PiEquation}
	\end{equation}
	where $b$ is constant, and
	\begin{align}
		\Pi\,(T)
			= & k_1 \, T + k_2 \, T \, e^{-1.76 T_c/T} \nonumber \\
			& \quad +1.76 \, k_2 \, T_c \, E_i
				\left(-1.76 \frac{T_c}{T}\right),
	\end{align}
	with $E_i(x) \equiv \int_{-x}^\infty [e^{-t} / t ]dt$ denoting the exponential integral function.
	$\zeta (r)$ is then the solution of Eq.~\eqref{eq:PiEquation} for $T$.
	Note that our simple model assumes that the quasiparticles and
	the phonons remain at the same effective temperature (the inelastic
	electronic mean free path is small), and that both remain diffusive
	(the elastic phonon and electron mean free paths are small). Violating
	either of these assumptions would likely lower the transport of 
	energy away from the grain boundary, making the heating more
	dangerous.
	
\begin{figure}[!ht]
		\centering
		\includegraphics[width=\linewidth]{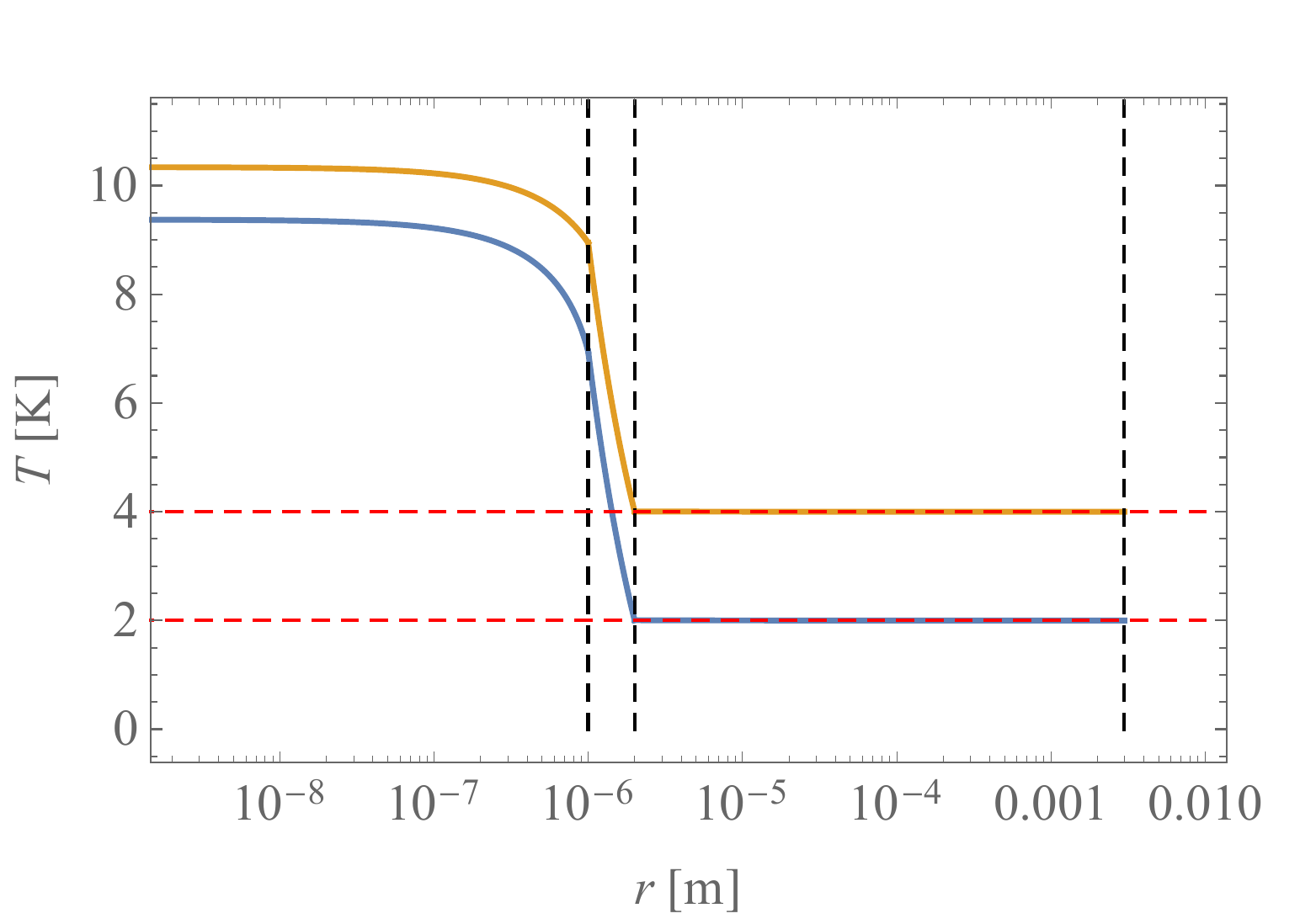}
		\caption{Temperature as a function of distance from the grain boundary for the
		Nb$_3$Sn/Nb system at a bath of 2K (blue) and 4K (yellow).
		Vertical dashed lines correspond to $D=1\mu$m, $R_1=2 \mu$m and $R_2=3$mm,
		from left to right.
		Bottom and top red dashed lines correspond to $T_\text{He}=2$K and 4K, respectively.
		We have used temperature-dependent $\kappaNbSb$ given by
		Eq.~\eqref{eq:ThermalConductivity}, $\kappaNb=10$W/m$\cdot$K, and $P_{GB}=621$nW
		at $B=60$mT, according to our previous estimates.}
		\label{fig:TPro}
	\end{figure}

	We focus our attention on grain-boundary activation and ignore other sources of power dissipation.
	These sources, in particular the BCS dissipation that exponentially increases with
	temperature, may have a significant impact on stationary temperature profiles, with possible
	indicatives of thermal runaway and cavity quenches at low fields.
	As we discuss at the end of this section, our simple estimates suggest that larger grain
	boundaries or multiple nearby boundaries could already raise the temperature high enough to
	quench the cavity.
	Hence, although the inclusion of other sources would lead to a more accurate description of
	heat diffusion near active grain boundaries, they will not lead to qualitative changes in our
	conclusions.
	Thus, we assume that the heat flux is the power dissipated per grain boundary ($P_{GB}$)
	per unit area.
	We use Fourier's law to determine the coefficients $a$, $c$ and $e$ in Eqs.~\eqref{eq:TSol}
	and~\eqref{eq:PiEquation}.
	For $0 \leq r \leq D$, $\dot{\mathcal{Q}} = P_{GB} / D^2$, so that:
	\begin{equation}
		a
			= \frac{P_{GB}}{D^2}.
	\end{equation}
	For $D \leq r \leq R_1$ ($R_1 \leq r \leq R_2$), $\dot{\mathcal{Q}} = P_{GB} / 2\pi r^2$,
	$\kappa=\kappaNbSb$ ($\kappaNb$) and $dT / dr = -c/r^2$ ($-e/r^2$), so that:
	\begin{equation}
		c
			= \frac{P_{GB}}{2 \pi \kappaNbSb}, \quad
		e
			= \frac{P_{GB}}{2 \pi \kappaNb}.
	\end{equation}
	To find $g$, we use $T (R_2) = T_\text{He}$, where $T_\text{He}$ is the temperature of the
	Helium bath:
	\begin{equation}
		g
			= T_\text{He} - \frac{P_{GB}}{2 \pi \kappaNb R_2}. \quad
	\end{equation}
	To find $d$ and $b$, we use the continuity of $T(r)$ at $r=R_1$ and $r=D$, respectively
	(we ignore the Kapitza resistance at the interface between Nb$_3$Sn and Nb).
	Thus,
	\begin{equation}
		c / R_1 + d
			= e/R_1 +g \quad \Rightarrow \quad
		d
			= \frac{e-c}{R_1} + g, 
	\end{equation}
	and
	\begin{equation}
		b
			= a D + \Pi \left(\frac{c}{D} + d \right).
	\end{equation}
	Figure~\ref{fig:TPro} shows temperature profiles (Eq.~\eqref{eq:TSol}) near an active grain
	boundary for Nb$_3$Sn/Nb systems at $B=60$mT.
	Note that the temperature of an active grain boundary increases to about 10K near the
	boundary surface for a Helium temperature of 2K.
	Although this increase in temperature is not large enough to drive Nb$_3$Sn into the
	normal state, it certainly has significant impact on the superconducting properties.
	Also, note that the temperature decays to nearly $T_\text{He}$ as $r$ approaches twice
	the grain size $D$, suggesting that heating due to grain-boundary activation is mostly
	localized.

	A temperature rise of 10K at the grain boundary is over half of
	the critical temperature of the film, suggesting that larger grain
	boundaries or multiple nearby boundaries could raise the temperature
	high enough to quench the cavity. Cavities with 
	tin-rich grain boundaries and more pristine grain boundaries
	show the same quench fields, suggesting that another mechanism
	controls the quench fields of existing Nb$_3$Sn cavities. 
	If the excess dissipation in the cavities with tin-rich boundaries
	is due to vortex penetration (Fig.~\ref{fig:QxVsB}), one would 
	expect rare events with large or multiple grain boundaries would
	happen, suggesting that our grain-boundary heating estimate is
	unduly pessimistic. Alternatively, it remains possible that the
	grain boundaries have high superheating fields, and the excess
	dissipation has another explanation. In any case, our estimates
	suggest that vortex entry at grain boundaries should be expected
	for tin-rich boundaries well below the superheating field for
	a perfect crystal, and that the subsequent heat release should be
	important both as a contribution to the overall dissipation 
	and as a quench mechanism for the cavity.

\section{Conclusion}
\label{sec:conclusion}

In this work we have presented an interdisciplinary, multi-scale study of vortex nucleation in Sn-segregated grain boundaries and its subsequent effect on SRF performance.  
Scanning transmission electron microscopy images and energy dispersive spectroscopy show Sn concentration as high as $\sim$35 at.\% and widths $\sim$3nm in chemical composition in grain boundaries.
We used density functional theory to estimate the effective critical temperature for the material in the segregation zone and find that the effective $T_c$ can be reduced to as low as 5 K for Sn concentrations in excess of $\sim$30 at.\%.
Next, we used these calculations as inputs into time-dependent Ginzburg-Landau simulations.
These simulations demonstrate that grain boundaries can act as nucleation sites for magnetic vortices.
The grain boundaries then act as pinning ``planes'' after nucleation that allow vortices to move vertically along the grain boundary, but prevent them from moving laterally into the bulk.
We have seen that for a range of applied fields, vortices may nucleate at but remain constrained to the grain boundary.
These vortices will nucleate and annihilate once per RF cycle, and we estimate the superconducting losses of this phenomenon at the scale of SRF cavities.
We have shown that as long as vortices do not penetrate the bulk grain, losses are localized near the grain boundary and will not lead to a global quench.
However, the annihilation process each cycle will lead to a reduction in the quality factor that increases with larger applied fields, consistent with the experimentally observed Q-slope.

SRF cavities are an important application area that require multi-disciplinary talents to address.
This study has leveraged the skills of accelerator physicists, material scientists, and condensed matter theorists with expertise across a range of scales to explore a question fundamental to the advancement of next-generation SRF material, Nb$_3$Sn.
This study has presented evidence that segregation zones in grain boundaries play an important role in cavity performance.
Understanding the mechanism behind the Q-slope will motivate new manufacturing protocols and help constrain the design space of future cavities.

\section{Acknowledgement}

We would like to thank Matthias Liepe, and Richard Hennig for helpful conversations.
This research is supported by the United States Department of Energy, Offices of High Energy. Fermilab is operated by the Fermi Research Alliance LLC under Contract No. DE-AC02-07CH11359 with the United States Department of Energy. This work made use of the EPIC, Keck-II, and/or SPID facilities of Northwestern University’s NUANCE Center, which received support from the Soft and Hybrid Nanotechnology Experimental (SHyNE) Resource (NSF ECCS-1542205); the MRSEC program (NSF DMR-1121262) at the Materials Research Center; the International Institute for Nanotechnology (IIN); the Keck Foundation; and the State of Illinois, through the IIN.
This work was supported by the US National Science Foundation under Award OIA-1549132, the Center for Bright Beams.

\bibliography{Bib.bib,GBEstimates.bib}

\end{document}